\documentclass[%
 reprint,
 aps,floatfix
]{revtex4-2}
\usepackage{tabularx}
\usepackage{amsmath}
\usepackage{amssymb}
\usepackage{amsfonts}
\usepackage{boxedminipage}
\usepackage{appendix}
\usepackage{algorithm2e}
\usepackage{mathtools}
\usepackage{physics}
\RestyleAlgo{boxruled}
\usepackage[utf8]{inputenc}

\usepackage[font=small]{caption}
\usepackage[labelformat=empty, position=top]{subcaption}
\usepackage{graphicx}
\usepackage[export]{adjustbox}

\usepackage{csquotes}
\MakeOuterQuote{"}

\usepackage{hyperref} 
 \hypersetup{%
 	colorlinks=true,       
 	linkcolor=blue,          
 	citecolor=blue,        
 	filecolor=magenta,      
 	urlcolor=blue         
 }
\usepackage{floatrow}
\usepackage{lmodern}

\newcommand{\Tee}{\mathsf{T}}

\newcommand{\includegraphicsTwoDown}[2]{
  \begin{subfigure}[t]{0.05\textwidth}
    \textbf{(a)}
  \end{subfigure}
  \begin{subfigure}[t]{0.93\textwidth}
    \includegraphics[width=\linewidth, valign=t]{#1}
  \end{subfigure}\hfill \\ \vspace{3mm}
  \begin{subfigure}[t]{0.05\textwidth}
    \textbf{(b)}
  \end{subfigure}
  \begin{subfigure}[t]{0.93\textwidth}
    \includegraphics[width=\linewidth, valign=t]{#2}
  \end{subfigure}
}

\newcommand{\includegraphicsTwoAcross}[2]{
  \begin{subfigure}[t]{0.05\textwidth}
    \textbf{(a)}
  \end{subfigure}
  \begin{subfigure}[t]{0.43\textwidth}
    \includegraphics[width=\linewidth, valign=t]{#1}
  \end{subfigure}
  \begin{subfigure}[t]{0.05\textwidth}
    \textbf{(b)}
  \end{subfigure}
  \begin{subfigure}[t]{0.43\textwidth}
    \includegraphics[width=\linewidth, valign=t]{#2}
  \end{subfigure}
}

\newcommand{\includegraphicsTwoByTwo}[4]{
  \begin{subfigure}[t]{0.05\textwidth}
    \textbf{(a)}
  \end{subfigure}
  \begin{subfigure}[t]{0.43\textwidth}
    \includegraphics[width=\linewidth, valign=t]{#1}
  \end{subfigure}
  \begin{subfigure}[t]{0.05\textwidth}
    \textbf{(b)}
  \end{subfigure}
  \begin{subfigure}[t]{0.43\textwidth}
    \includegraphics[width=\linewidth, valign=t]{#2}
    \hfill \\ \vspace{3mm}
    \end{subfigure}
  \begin{subfigure}[t]{0.05\textwidth}
    \textbf{(c)}
  \end{subfigure}
  \begin{subfigure}[t]{0.43\textwidth}
    \includegraphics[width=\linewidth, valign=t]{#3}
  \end{subfigure}
  \begin{subfigure}[t]{0.05\textwidth}
    \textbf{(d)}
  \end{subfigure}
  \begin{subfigure}[t]{0.43\textwidth}
    \includegraphics[width=\linewidth, valign=t]{#4}    
  \end{subfigure}
}

\begin{document}
\title{A Reduction-Based Strategy for Optimal Control of Bose-Einstein Condensates}
\author{J. Adriazola}
\author{R. H. Goodman}
\affiliation{Department of Mathematical Sciences, New Jersey Institute of Technology, University Heights, Newark, NJ 07102}
\date{\today}
\begin{abstract}
Applications of Bose-Einstein Condensates (BEC) often require that the condensate be prepared in a specific complex state. Optimal control is a reliable framework to prepare such a state while avoiding undesirable excitations, and, when applied to the time-dependent Gross-Pitaevskii Equation (GPE) model of BEC in multiple space dimensions, results in a large computational problem. We propose a control method based on first reducing the problem, using a Galerkin expansion, from a PDE to a low-dimensional Hamiltonian ODE system. We then apply a two-stage hybrid control strategy. At the first stage, we approximate the control using a second Galerkin-like method known as CRAB to derive a finite-dimensional nonlinear programming problem, which we solve with a differential evolution (DE) algorithm. This search method then yields a candidate local minimum which we further refine using a variant of gradient descent. This hybrid strategy allows us to greatly reduce excitations both in the reduced model and the full GPE system. 
\end{abstract}

\maketitle

\section{\label{sec:intro}Introduction and Experimental Context}
Quantum optimal control is concerned with the control of $N$-body quantum systems~\cite{glaser2015training,BorziBook}. One important example is the reshaping of a dilute atomic Bose-Einstein condensate (BEC). Since they were observed in laboratory experiments in 1995~\cite{Weiman, PhysRevLett.75.1687,PhysRevLett.75.3969}, BECs, an ultra-cold quantum fluid whose mean dynamics resemble that of a single atom~\cite{GPEref}, have proven to be an experimentally reliable and versatile platform for high-precision quantum metrology~\cite{gross2010, lucke2011, riedel2010}, high-precision storage, manipulation, and probing of interacting quantum fields~\cite{RevModPhys.80.885, ockeloen2013}. Future   quantum computation and simulation  technologies will likely require fast manipulation of BECs~\cite{calarco2000, kielpinski2002}. 

Experimentalists over the last two decades have achieved remarkably high, yet sub optimal degrees of control of BECs by using empirical rules of thumb and intuition gained from significantly reduced models admitting closed-formed solutions~\cite{Nature}. Meanwhile, optimal control theory provides a computational framework for systematically finding highly efficient control policies~\cite{peirce1988, koch2004, kirk2004optimal, brif2010control}. The success of optimal control theory is demonstrated numerically in three spatial dimensions by Mennemann, et al.~\cite{Mennemann}. Our work interpolates between these two approaches by applying a general optimization strategy to simpler, ordinary differential equations (ODE) which are, in some sense, still faithful to the  partial differential equations (PDE) that model BEC.

\begin{figure}[tb]
  
  \includegraphicsTwoAcross{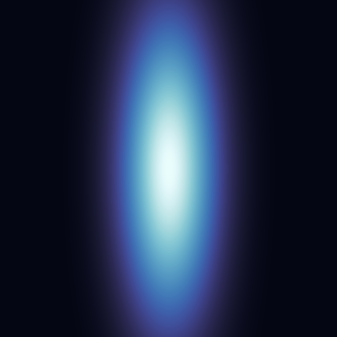}{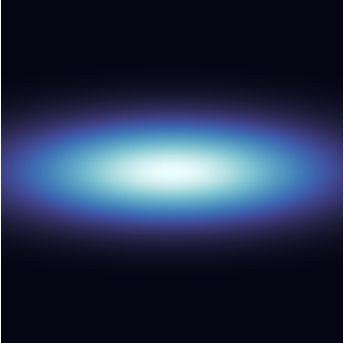}
  \caption{Reshaping a BEC. \textbf{(a)}~The density distributions $\abs{\psi_{\rm Gauss}}^2$ axially aligned along the vertical with $(a,b)=(10,1)$ in Equation~\eqref{eq:introsqueezewave}. \textbf{(b)}~The density distribution axially aligned along the horizontal with $(a,b)=(1,10)$.}
  \label{fig:MenRot}
\end{figure}

\begin{figure}[tb]
  
  \includegraphicsTwoAcross{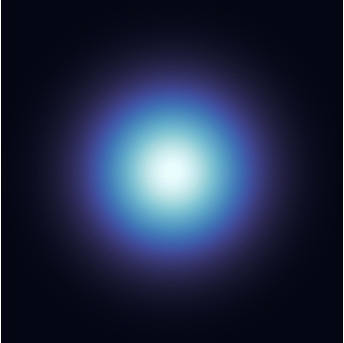}{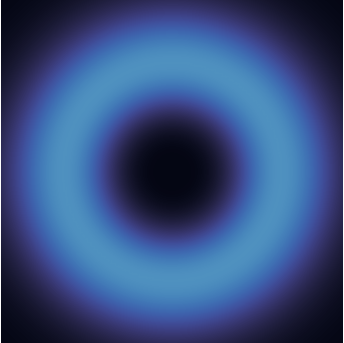}
  \caption{Topologically changing the condensate's support. \textbf{(a)}~The initial distribution $\abs{\psi_{\rm Gaussian}}^2$ with $a=b=1$. \textbf{(b)}~The desired distribution $\abs{\psi_{\rm Toroid}}^2$.}
  \label{fig:introsplit}
\end{figure}

Mennemann, et al., apply optimal control to reshape the support of a BEC, reorienting the magnetic field  concentrated along one axial direction to an direction, which in turn reorients the density distribution of the condensate.
A two dimensional schematic is shown in Figure~\ref{fig:MenRot} with normalized Gaussian wavefunctions of the form 
\begin{equation}\label{eq:introsqueezewave}
    \psi_{\rm Gauss}=\sqrt[4]{\frac{ab}{\pi^2}}e^{-ax^2-by^2}, 
\end{equation}
where $a,b>0$ and $(x,y)\in\mathbb{R}^2$. Another manipulation Mennemann, et al., consider is to change the topology of the wavefunction's support. An example of this is shown in Figure~\ref{fig:introsplit} where a Gaussian wavefunction~\eqref{eq:introsqueezewave}, with $a=b=1,$ is mapped to the normalized toroidal wavefunction
$$
    \psi_{\rm Toroid}=\frac{2}{\sqrt{3\pi}}r^2e^{-r^2}, \text{ where } r^2 = x^2 + y^2.
$$
Manipulating the condensate excites oscillations that prevent the transformed distribution from matching the desired distribution after the control process has terminated. In this paper we use optimal control to perform these manipulations while minimizing such oscillations and mismatch.

The, now standard, optimal control problem, first proposed by Hohenester, et al.~\cite{Hohenester}, is to maximize the fidelity between an evolving field $\psi$ at a final time $T>0$ and an experimentally-desired state $\psi_d$, subject to a control function $u$.  This problem, expressed in dimensionless form, is
\begin{equation}
\label{eq:HohenObj}
\inf_{u\in \mathcal{U}}{J}=\frac{1}{2}\inf_{u\in \mathcal{U}} \left [J^{\rm infidelity}(u) + J^{\rm regular}(u)\right], 
\end{equation}
where
\begin{equation}\begin{split}
J^{\rm infidelity}(u)&= 1-\abs{\left\langle\psi_d(x),\psi(x,T)\right\rangle}^2_{L^2(\mathbb{R}^3)} \qand \\
J^{\rm regular}(u)&=\gamma\int_0^T|\dot{u}|^2 \dd t,
\label{eq:Jdef}
\end{split}
\end{equation}
subject to 
\begin{subequations}
\begin{align}
\label{eq:Intro1}
i\partial_t\psi+\frac{1}{2}\nabla^2\psi-V(x,u(t))\psi-|\psi|^2\psi=0&, \\
\psi(x,0)=\psi_0(x)\in H^{1}\left(\mathbb{R}^3\right)&,\\
\|\psi(x,\cdot)\|_{L^2(\mathbb{R}^3)}=1&,
\end{align}
\end{subequations}
where $\gamma>0,\ t\in[0,T],$ the wavefunction $\psi(x,t)$ belongs to $L^2\left([0,T];H^{1}\left(\mathbb{R}^3\right)\right)$, $\psi_0$ is some initial state,  $\nabla^2$ is the Laplacian operator, $V(x,u)$ models the geometry of confinement as a potential energy parameterized by the control $u(t)$,  $\mathcal{U}$ is an admissible class of  control functions $u\in H^1([0,T])$ with fixed initial and terminal conditions, $L^2(\Omega)$ is the space of square Lebesgue-integrable functions over the measurable set $\Omega$, and $H^{1}(\Omega)$ is the Sobolev space of $L^2(\Omega)$ functions whose first weak derivatives are also in $L^2(\Omega)$.

The dynamical constraint~\eqref{eq:Intro1} is known as the Gross-Pitaevskii equation (GPE), and the confining potential $V(x,u(t))$ arises due to an applied optical or magnetic field. How the constraint arises as a model of the mean-field dynamics of BEC is discussed in detail in~\cite{GPEref}. The terms $J^{\rm infidelity}$ and $J^{\rm regular}$ are known as the infidelity and regularization terms. In the language of optimal control theory~\cite{Bryson,Calculus1989}, the infidelity is a type of \emph{terminal cost} which penalizes control policies that miss the desired wave function $\psi_d$. The regularization is a type of \emph{running cost} which penalizes the usage of physically undesirable controls with fast variations and ensures that Hohenester's optimal control problem remains well-posed. This is shown for a more general control problem defined by the Hohenester objective~\eqref{eq:HohenObj} and mean field constraint~\eqref{eq:Intro1}, along with a running cost which also penalizes the amount of work done by the control, in work due to Hintermuller, et al.~\cite{Hintermuller}. 

Mennemann, et al., numerically study experimentally motivated transformations of $\psi(x,t),$ such as those in Figures~\ref{fig:MenRot} and~\ref{fig:introsplit}, by solving the associated optimal control problem, after setting $\gamma=10^{-6},$ with a projected gradient method called Gradient Pulse Engineering (GRAPE). Their work is the source of inspiration for this paper.

We begin with two primary questions: can we gain further physical intuition of the condensate dynamics as it is controlled, and can we use this physical insight to implement optimization strategies in some easier, i.e., finite-dimensional, computational setting? To this end, we introduce a Galerkin ansatz which incorporates the time-dependence of the confining potential, and use this to study two model problems in one space dimension: squeezing a BEC in a steepening quadratic potential and splitting a BEC with a time-dependent barrier. These model problems abstract the salient features of the reorientation problem illustrated in Figure~\ref{fig:MenRot} and the splitting problem illustrated in Figure~\ref{fig:introsplit}.

\subsection{Structure of the Paper}
In Section~\ref{sec:Modeling} we describe the Galerkin reduction of the squeezing and splitting problems, following, for example~\cite{JPhysA2011,gmw_2015}. We assume the reshaping potential is product-separable in space and time, the problem is even-symmetric, and initial conditions are small-amplitude superpositions of eigenfunctions of the associated linear Schrödinger operator so that the dynamics are weakly nonlinear. Using these assumptions, we reduce the dynamics of the controlled condensate  a non-autonomous one degree of freedom Hamiltonian system using Galerkin reductions and canonical transformations. We validate this reduction by comparing numerical solutions of the GPE with a specified time-dependent potential with solutions of the reduced system.

We then pose an optimal control problem, in Section~\ref{sec:BECoptimization}, constrained by the Hamiltonian dynamics whose objective to minimize that same Hamiltonian function and thereby minimize oscillations which persist after the control process is terminated. We then provide the necessary optimality conditions for this class of Hamiltonian control problems.

In Section~\ref{sec:numerics}, we outline the numerical methods used to solve the nonconvex optimization problem of Section~\ref{sec:BECoptimization}. We proceed in two steps, using a so-called hybrid method. Because the objective function is non-convex, it may have many local minima, and we first must search for the best among many candidates. We use a method due to Calarco, et al.~\cite{Doria,Caneva}, called the Chopped Random Basis (CRAB) method to reduce the search space to finite dimensions by considering controls within a space of Galerkin approximations. This space is searched using a global, nonconvex method due to Storn and Price called Differential Evolution (DE)~\cite{Storn}. The second step is to refine the best candidate using a local descent method. The GRAPE method allows us to perform this descent among controls satisfying desired boundary conditions.

To validate the proposed approach, in Section~\ref{sec:Results} we solve the Gross-Pitaevskii equation~\eqref{eq:Intro1} again, using the controls resulting from the methods of Section~\ref{sec:numerics}. We find that this approach both suppresses undesirable persistent  oscillations and minimizes the infidelity in the Hohenester objective functional~\eqref{eq:HohenObj}. 

\section{\label{sec:Modeling}Model Problems from a Galerkin Reduction}

In this section, we outline the derivation of model Hamiltonian problems via a Galerkin truncation. We apply this truncation to a GPE in one spatial dimension which we assume depends on a stationary potential $V_s(x)$ and a reshaping potential $V_r(x)$, i.e.,
\begin{equation}\label{eq:GPE}
    i\partial_t\psi=-\frac{1}{2}\partial_x^2\psi+V_s(x)\psi+u(t)V_r(x)\psi+|\psi|^2\psi.
\end{equation}
We use a Galerkin expansion of the form
\begin{equation}\label{eq:GPEGal}
    \psi(x,t)=\sum_{n=0}^{\infty}c_n(t)\varphi_n(x;u(t)),
\end{equation}
where each of the basis functions $\varphi_n(x;u(t))$ is an instantaneously normalized eigenfunction of the equation
\begin{equation*}
    -\frac{1}{2}\partial_x^2\varphi_n+\left(V_s(x)+uV_r(x)\right)\varphi_n=E_n\varphi_n,
\end{equation*}
i.e., the linear Schr\"{o}dinger equation with $u$-dependent potential. 

By choosing initial conditions which are in the form of~\eqref{eq:GPEGal} with $\abs{c_n }$ relatively small, nonlinear effects remain relatively weak throughout the control process. This allows us to truncate the expansion~\eqref{eq:GPEGal} at a low order. This large reduction of dimension due to the Galerkin greatly simplifies the dynamics and is justified through numerical studies in Section~\ref{sec:galerkin_experiment}. We show the coefficients $c_n(t)$ evolve under a Hamiltonian system whose dynamics motivates the control strategy discussed in Section~\ref{sec:BECoptimization}.

\subsection{The Squeezing Problem}\label{sec:Squeeze}

We first address the problem of squeezing and elongation discussed in Section~\ref{sec:intro} and shown in Figure~\ref{fig:MenRot}. As a model problem, we consider the squeezing of a stationary wave packet centered about the origin and trapped in a reshaping quadratic potential, i.e. $V_r(x)=\frac{1}{2}x^2,\ V_s(x)\equiv0$ in Equation~\eqref{eq:GPE}, with the endpoints of the control fixed as $u(0)=u_0>0$ and  $u(T)=u_T>u_0$. 

In this case, each of the $\varphi_n(x;u(t))$ in Expansion~\eqref{eq:GPEGal} satisfies
\begin{equation*}
    -\frac{1}{2}\partial_x^2\varphi_n+\frac{1}{2}ux^2\varphi_n=E_n\varphi_n.
\end{equation*}
The eigenfunctions $\varphi_n(x;u)$ are the well-known Hermite functions and can be generated by the Rodrigues formula
\begin{equation}\label{eq:hermitebasis}
    \varphi_n(x;u)= (-1)^n\frac{\pi^{-1/4}}{\sqrt{2^nu^{n/4}\,n!}}u^{1/8}e^{\frac{u^{1/4}x^2}{2}}\partial_x^ne^{-u^{1/4}x^2}.
\end{equation} 
The first three are
\begin{align*}
\varphi_0(x;u)&= \xi e^{-\frac{1}{2} \sqrt{u} x^2},\\
\varphi_1(x;u)&=\sqrt{2}\xi u x e^{-\frac{1}{2} \sqrt{u} x^2},\\
\varphi_2(x;u)&=\sqrt{2}\xi u\left(2 \sqrt{u} x^2-1\right)e^{-\frac{1}{2} \sqrt{u} x^2}, 
\end{align*}
where $\xi=\pi^{-1/4}u^{1/8}$.
We truncate expansion~\eqref{eq:GPEGal} after the third term, and discard the single odd term involving $\varphi_1(x;u)$ because we assume the initial conditions obey an even symmetry which is invariant under GPE. For convenience, we relabel these first two even eigenstates and their time dependent coefficients as the $n=0,1$ states.  

To derive the equations governing the time-dependent coefficients present in expansion~\eqref{eq:GPEGal}, we substitute the expansion into the GPE~\eqref{eq:GPE} and project onto each mode using the standard $L^2\left(\mathbb{R}\right)$ inner product. Letting $\dagger$ denote complex conjugation and overhead dots denote time derivatives, the resulting ODE system is Hamiltonian, i.e., 
\begin{equation}\label{eq:Hamiltons}
    i\dot{c}_n=\partial_{c^{\dagger}_n}\mathcal{H},\qquad i\dot{c}^{\dagger}_n=-\partial_{c_n}\mathcal{H},\qquad n=0,1,
\end{equation}
with the Hamiltonian $\mathcal{H}\left(c_0,c_0^{\dagger},c_1,c_1^{\dagger};u\right)$ given by
\begin{align}\label{eq:SqueezHam}
    \mathcal{H}=\xi^2
    \left(
    \frac{\abs{c_0}^4}{2 \sqrt{2  }}
    +\frac{41 \abs{c_1}^4}{128 \sqrt{2  }}
    +\frac{3\abs{c_0}^2 \abs{c_1}^2}{4 \sqrt{2  }}   
    +\frac{3\Re\left\{c_0^2 c^{\dagger 2}_1\right\} }{8 \sqrt{2  }}\right. \nonumber\\
    \left.
    -2\Re\left\{c_0 c^{\dagger}_1\right\}\left(
    \abs{c_0}^2-\frac{\abs{c_1}^2}{8}\right)
    \right) \nonumber\\
+\frac{\sqrt{u}}{2} \left(\abs{c_0}^2 +5 \abs{c_1}^2\right)-\frac{\dot{u}}{2 \sqrt{2} u}\Im\left\{c_0c^{\dagger}_1\right\}.
\end{align}
Note that, the dynamics conserve the "discrete" mass
\begin{equation}\label{eq:discretemass}
    M_d(t)=\abs{c_0(t)}^2+\abs{c_1(t)}^2.
\end{equation} 

Next, we reduce the the squeezing Hamiltonian~\eqref{eq:SqueezHam} to one and a half degrees of freedom using canonical transformations. This allows the use of phase plane techniques which provide further insight into the problem.

We first convert to action-angle coordinates through the canonical transformation
\begin{equation*}
c_0=\sqrt{\rho_0}e^{-i\theta_0},\quad c_1=\sqrt{\rho_1}e^{-i\theta_1}.
\end{equation*}
Hamiltonian~\eqref{eq:SqueezHam} then becomes
\begin{align*}
\mathcal{H}=&\frac{\sqrt{u}}{2}\left(\rho _0+5 \rho _1\right)-\frac{\dot{u}}{4u} \sqrt{2\rho _0\rho _1}\sin \phi+\\
&\frac{3\xi^2}{8\sqrt{2}} \rho _0 \rho _1  \cos (2 \phi )
+\frac{\sqrt{2}\xi^2}{256}\left(64  \rho _0^2+96  \rho _1 \rho _0+\right. \nonumber\\
    &\left.41  \rho _1^2+\left(8 \rho _1^{3/2}\sqrt{2\rho _0}-56 \rho _0^{3/2} \sqrt{2\rho _1}\right) \cos (\phi)\right),
\end{align*}
where $\phi=\theta_0-\theta_1$. In these coordinates, the discrete mass~\eqref{eq:discretemass} is given by $M_d=\rho_0+\rho_1$. We make the choice to set the discrete mass to one, and introduce the change of variables $\rho_0=1-J$ and $ \rho_1=J$ so that
\begin{align*}
\mathcal{H}=\frac{\xi^2}{128\sqrt{2}}\left(9 J^2-32J+64\right)-\frac{\dot{u}}{2 \sqrt{2} u}\sqrt{(1-J) J} \sin (\phi )\nonumber\\
+\frac{\sqrt{u}}{2} \left(1+4J\right)
+\frac{\xi^2}{16}\left( \sqrt{(1-J) J}(8-7 J) \cos (\phi )\right. \nonumber\\
    \left.+3\sqrt{2} (1-J) J\cos (2 \phi )\right).
\end{align*}
In these coordinates, $J=0$ indicates that all of the mass is in the ground state while $J=1$ indicates that all of the mass is in the excited state.

A further canonical transformations facilitate visualization of the phase portrait. Defining $q+i p =\sqrt{2J}e^{i\phi}$ yields
\begin{align}\label{eq:FinalSqueezeHam}
&\mathcal{H}(q,p,u)=\sqrt{u}\left(q^2+p^2+\frac{1}{2}\right)
\nonumber\\&+\frac{\xi^2}{64}\sqrt{2-p^2-q^2}
\left(9 q^3 
    -16 q+ 9 p^2 q -\frac{8 \sqrt{2 \pi }\dot{u}}{u} p\right)\nonumber\\
&+\frac{\xi^2}{512\sqrt{2}}\left(57 p^4-160 p^2+18 p^2 q^2-39 q^4+32 q^2+256\right).
\end{align}

\subsection{The Splitting Problem}\label{sec:Split}
We refer to the problem of topologically changing the support of the condensate, mentioned in Section~\ref{sec:intro} and shown in Figure~\ref{fig:introsplit}, as the "splitting" problem. In the case of one spatial dimension and "splitting" potential $V_r(x)=\delta(x),$ together with quadratic stationary potential $V_s(x)=\frac{1}{2}x^2,$  the linear Schrödinger equation is exactly solvable for each value of $u$. 

In order to construct the Galerkin ansatz, we provide brief details on solving the eigenvalue problem
\begin{equation}\label{eq:SplitEigen}
    -\frac{1}{2}\partial_x^2\varphi_n+\frac{1}{2}x^2\varphi_n+u\delta(x)\varphi_n=E_n\varphi_n.
\end{equation}
A more thorough computation and discussion is given by Viana-Gomes and Peres~\cite{Viana-Gomes2011}. First, note that integrating~\eqref{eq:SplitEigen} in a neighbourhood about the origin leads to a jump condition on the derivative:
\begin{equation}\label{eq:splitcond}
    \lim_{\varepsilon\to0}\partial_x\varphi(x)\big|_{-\varepsilon}^{+\varepsilon}=2u\varphi\big|_{x=0}.
\end{equation}
Since all odd $C^1(\mathbb{R})$ functions satisfy the jump condition, the odd-parity states are given by the Rodrigues formula~\eqref{eq:hermitebasis} with $u=1$. Only the even-parity states are modified by the delta function at the origin. 

By letting $\varphi=e^{-x^2/2}w(x),$  $z=x^2$, and  $E=\nu+\frac{1}{2},\ \nu\in\mathbb{R},$ Equation~\eqref{eq:SplitEigen} and condition~\eqref{eq:splitcond} become
\begin{subequations}
\begin{align}\label{eq:KummerEq}
    z\partial_z^2w+\left(\frac{1}{2}-z\right)\partial_zw+\frac{\nu^2}{2}w&=0,\quad z>0,\\ 
    \label{eq:evensplitcond}
    \partial_zw\big|_{z=0}=uw\big|_{z=0}.
\end{align}
\end{subequations}
Equation~\eqref{eq:KummerEq} is called Kummer's equation and admits solutions of the form
\begin{equation*}
    w(z)=A_{\nu}U\left(-\frac{\nu}{2},\frac{1}{2},z\right),
\end{equation*}
where
\begin{align}\label{eq:Tricomi}
U\left(a,b,z\right)=\frac{\Gamma(1-b)}{\Gamma(a+1-b)}&M(a,b,z)\nonumber\\+\frac{\Gamma(b-1)}{\Gamma(a)}z^{(1-b)}&M(a+1-b,2-b,z),
\end{align}
is Tricomi's confluent hypergeometric function whose definition involves the  gamma function, $\Gamma(z),$ and Kummer's function
\begin{equation*}
M(a,b,z)=\sum_{n=0}^{\infty}\frac{a^{(n)}z^n}{b^{(n)}n!},
\end{equation*}
with $(\cdot)^{(n)}$ denoting the rising factorial defined by
\begin{equation*}
    a^{(n)}:=\prod_{k=0}^{n-1}(a+k).
\end{equation*}
The coefficient $A_{\nu}$ is a normalization constant.

Applying boundary condition~\eqref{eq:evensplitcond}
to $w(z)$ leads to the nonlinear equation
\begin{equation}\label{eq:nuimp}
\nu-u\frac{\Gamma\left(1-\frac{\nu}{2}\right)}{\Gamma\left(\frac{1}{2}-\frac{\nu}{2}\right)}=0
\end{equation}
for $\nu$. For $u$=0, Equation~\eqref{eq:nuimp} implies $\nu$=0, and we recover the even Hermite basis given by the Rodrigues formula~\eqref{eq:hermitebasis}. In general, a numerical solution of Equation~\eqref{eq:nuimp}, demonstrated in~\cite{Viana-Gomes2011}, shows there is a countable sequence of solutions $\{\nu_n\}$ each satisfying $\nu_{n+1}=\nu_n+2$. Thus it suffices to solve Equation~\eqref{eq:nuimp} on the interval [0,1], the interval containing the ground state value of $\nu$, since this determines all other solutions. Therefore, we m restriction $\nu$ to $[0,1],$ so that the first two even eigenfunctions can be written as
\begin{align}
\label{eq:trieig}
\varphi_j(x;\nu)&=N_j(\nu)e^{-\frac{x^2}{2}}U\left(-\frac{\nu+2j}{2},\frac{1}{2},x^2\right)\nonumber\\&:=N_j(\nu)e^{-\frac{x^2}{2}}U_j(x^2,\nu),\quad j=0,1,
\end{align}
where $N_j(\nu)$ are $\nu$-dependent normalization constants given by
\begin{equation*}
N_j^{-2}(\nu)=\int_{\mathbb{R}}e^{-x^2}U_j^2\left(x^2,\nu\right)\dd x,\quad j=0,1.
\end{equation*} 
These eigenfunctions $\varphi_j(x;\nu)$ will serve as basis functions in the Galerkin expansion. 

\begin{figure}[htbp]
  
  \includegraphicsTwoDown{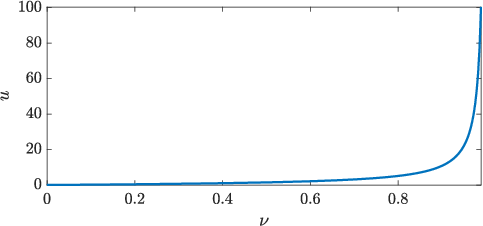}{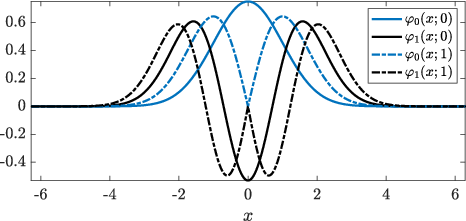}
  \caption{\textbf{(a)}~The function $u(\nu)$ as determined by Equation~\eqref{eq:nuimp}. Large values of $u$ are needed to achieve $\nu$ close to $1$. \textbf{(b)}~The first two even states in~\eqref{eq:trieig} for the values of $\nu=0,1$.}
 \label{fig:SplitModel}
\end{figure}

Note that as $\nu\to1,$ $u\to\infty$, since $\Gamma(z)$ has a pole at the origin. In this case, the first two even eigenfunctions reduce to the simple form of "split" wavefunctions
\begin{align*}
\varphi_0(x;1)&=2^{\frac{1}{2}}\pi^{-\frac{1}{4}}\abs{x}e^{-\frac{x^2}{2}},\\
\varphi_1(x;1)&=2\pi^{-\frac{1}{4}}3^{-\frac{1}{2}}\left(\abs{x}^3-\frac{3}{2}\abs{x}\right)e^{-\frac{x^2}{2}},
\end{align*}
shown in Figure~\ref{fig:SplitModel}.

Proceeding as Section~\ref{sec:Squeeze}, we have,
after projecting onto each mode using the $L^2\left(\mathbb{R}\right)$ inner product, a Hamiltonian system with Hamiltonian
\begin{align}\label{eq:SplitHam}
    \mathcal{H}=&\left(\alpha_0+u\beta_0\right)\abs{c_0}^2+\left(\alpha_2+u\beta_2\right)\abs{c_1}^2\nonumber\\
    &+2\left(\alpha_1+\beta_1\right)\Re\left\{c_0c_1^{\dagger}\right\}+\frac{1}{2}\gamma_0\abs{c_0}^4+\frac{1}{2}\gamma_1\abs{c_1}^4\nonumber\\
    &+2\left(\gamma_3\abs{c_0}^2+\gamma_2\abs{c_1}^2\right)\Re\left\{c_0c_1^{\dagger}\right\}\nonumber\\
    &+\gamma_4\left(\abs{c_0}^2\abs{c_1}^2+2\Re\left\{c_0^2c_1^{\dagger2}\right\}\right)+2\Delta\Im\left\{c_0^{\dagger}c_1\right\},
\end{align}
where the projection coefficients are given by
\begin{align} \label{eq:splitprojco}
&2\alpha_0=\left\langle\varphi_0,x^2\varphi_0-\partial_x^2\varphi_0\right\rangle,\ 2\alpha_1=\left\langle\varphi_0,x^2\varphi_1-\partial_x^2\varphi_1\right\rangle,\nonumber\\ &2\alpha_2=\left\langle\varphi_1,x^2\varphi_1-\partial_x^2\varphi_1\right\rangle,
\nonumber\\
&\beta_0=\left\langle\varphi_0,\delta(x)\varphi_0\right\rangle,\ \beta_1=\left\langle\varphi_0,\delta(x)\varphi_1\right\rangle,\nonumber\\ &\beta_2=\left\langle\varphi_1,\delta(x)\varphi_1\right\rangle,\nonumber\\
&\gamma_0=\|\varphi_0^4\|,\
\gamma_1=\|\varphi_1^4\|,\
\gamma_2=\left\langle\varphi_0,\varphi_1^3\right\rangle,\nonumber\\ &\gamma_3=\left\langle\varphi_1,\varphi_0^3\right\rangle,\
\gamma_4=\left\langle\varphi_0^2,\varphi_1^2\right\rangle,\nonumber\\
&\Delta=\left\langle\varphi_0,\partial_t\varphi_1\right\rangle=-\left\langle\partial_t\varphi_0,\varphi_1\right\rangle.
\end{align}
Applying the same canonical transformations as in Section~\ref{sec:Squeeze}, we arrive at
\begin{align}\label{eq:SplitHamqp}
    &\mathcal{H}=\alpha _0+\frac{\gamma _0}{2}+\frac{1}{2}(p^2+q^2) \left(\alpha _2-\alpha _0-\gamma _0+(\beta _2 -\beta _0) u\right)\nonumber\\
    &+\sqrt{2-p^2-q^2} \nonumber\\
    &\left(q \left( \alpha _1+\gamma _3+\frac{1}{2}\left(\gamma _2-\gamma _3\right) \left(p^2+q^2\right)+ \beta _1 u\right)+ \Delta  p\right)\nonumber\\
    &+\frac{\gamma_4}{2}\left(3q^2-p^2\right)
    +\frac{1}{8}\left(\gamma _0+\gamma _1\right)\left(p^4+q^4\right) \nonumber\\
    &+\frac{\gamma_4}{4}\left(p^4-3q^4\right)+\frac{1}{4} \left(\gamma _0+\gamma _1-2 \gamma _4\right) p^2 q^2+\beta _0 u.
\end{align}

\subsection{Numerical Experiments}
\label{sec:galerkin_experiment}

In this section, we simulate both the GPE~\eqref{eq:GPE} and the Galerkin-truncated systems of ODE describing both the squeezing and splitting problems, given by Hamiltonians~\eqref{eq:SqueezHam} and~\eqref{eq:SplitHam}, respectively, and compare the results. For both problems we fix the time domain $t\in[0,T],\ T>0,$ and solve both the GPE and the reduced models using a control of the form
\begin{equation}\label{eq:ramp}
    u(t)=(u_T-u_0)\frac{t}{T}+u_0,
\end{equation}
with $u_0=1,\ u_T=100,$ and $T=2.5$ for the squeezing experiment, and $u_0=0,\ u_T=30,$ and $T=10$ for the splitting experiment. 

In the ODE models we choose the initial conditions $(c_0(0),c_1(0))$ to minimize the associated Hamiltonian subject to the choice $M_d=1$, where $\dot{u}$ is set to zero in the definition of $\mathcal{H}$ in Hamiltonians~\eqref{eq:SqueezHam} and~\eqref{eq:SplitHam}. Thus the initial conditions are taken to be the fixed point of the system before the control is applied. The GPE is initialized as a superposition of the first two even states, 
\begin{equation*}
    \psi_0(x)=c_0(0)\varphi_0(x,u_0)+c_1(0)\varphi_1(x,u_0),
\end{equation*}
so that it represents the same initial state.

We solve the GPE~\eqref{eq:GPE} using a second-order-in-time split-step Fourier method using the midpoint method to integrate the time-dependence on the potential $u(t)V_r(x)$ and solve the ODE systems using MATLAB's \texttt{ode45}, i.e., an adaptive-step fourth order Runge-Kutta method. To compare the numerical solution of the GPE system with the numerical solutions to the  Galerkin truncated systems, we define the projected solution and the instantaneous Galerkin coefficients by projecting the numerical solution of GPE onto the instantaneous eigenfunctions, 
\begin{align*}
    \psi_{\rm proj}(x,t)&=\sum_{n=0}^1\left\langle\psi_{\rm GPE}(x,t),\varphi_n(x;u(t))\right\rangle\varphi_n(x;u(t))\nonumber\\&:=\sum_{n=0}^1c_{n}^{\rm proj}(t)\varphi_n(x;u(t)).
\end{align*}
We may also construct the approximate solution to GPE $\psi_{\rm galerkin}$ by evaluating the Galerkin ansatz using the numerically calculated values of $c_0(t)$ and $c_1(t)$.

\begin{figure*}[htbp]
  \includegraphicsTwoByTwo{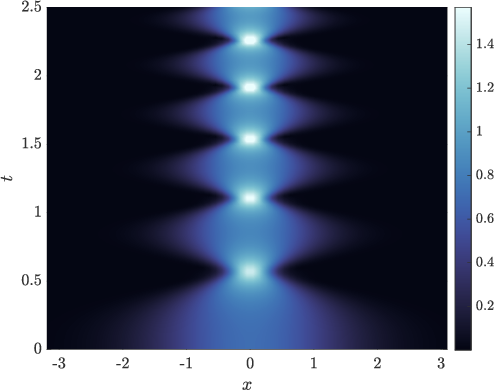}{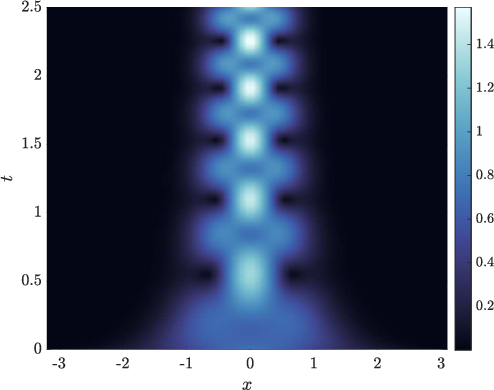}{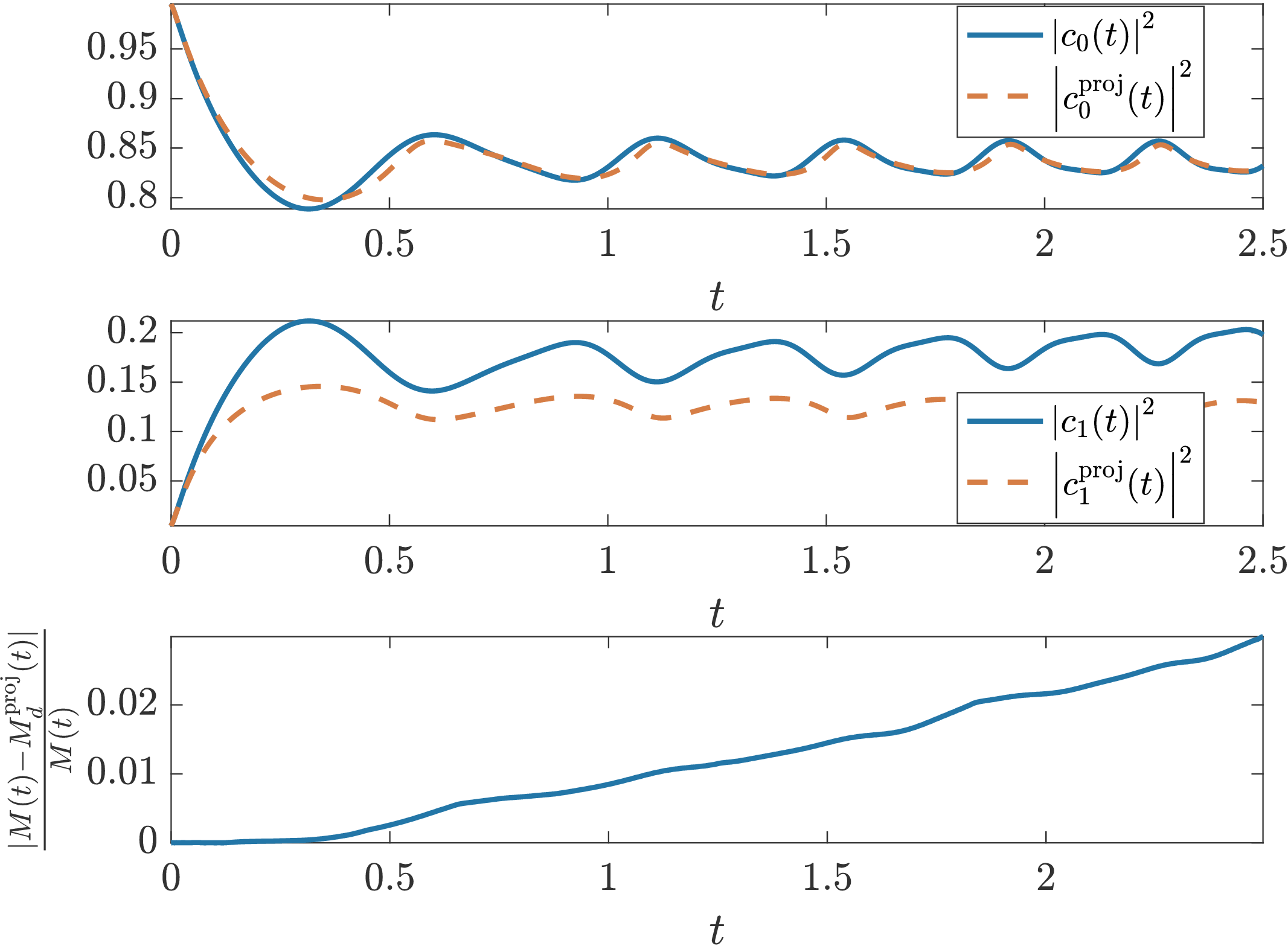}{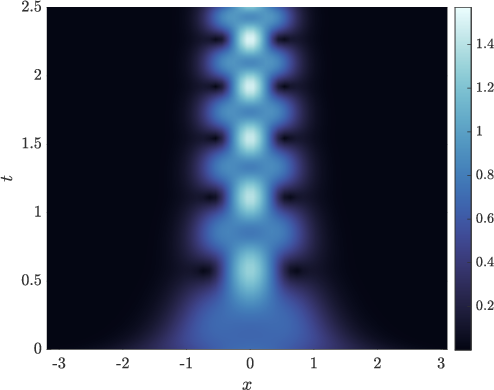}
\caption{\textbf{(a)}~Full numerical solution to Equation~\eqref{eq:GPE} using the  quadratic potential with time dependence~\eqref{eq:ramp}, with $\abs{ \psi }^2$ plotted versus $x$ and $t$. \textbf{(b)}~The Galerkin truncation of this solution to the first two even eigenfunctions. \textbf{(d)}~An approximate PDE solution constructed from an equivalent solution to ODE system~\eqref{eq:Hamiltons}. \textbf{(c)} First two rows: $\abs{c_0}^2$ and $\abs{c_1}^2$  computed by solving the ODE system (blue) and solving the PDE system and then projecting (red, dashed). Bottom row: Relative error  between the full PDE solution and the projection.}
\label{fig:BECSqueezTest}
\end{figure*}

\begin{figure*}[htbp]
\includegraphicsTwoByTwo{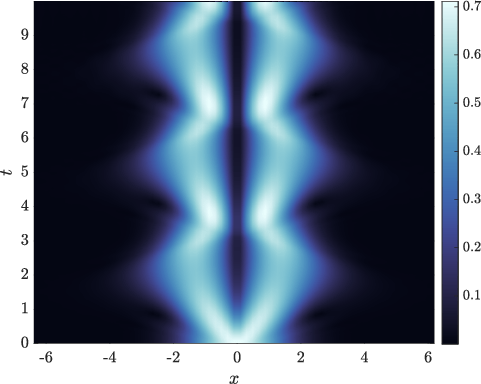}
{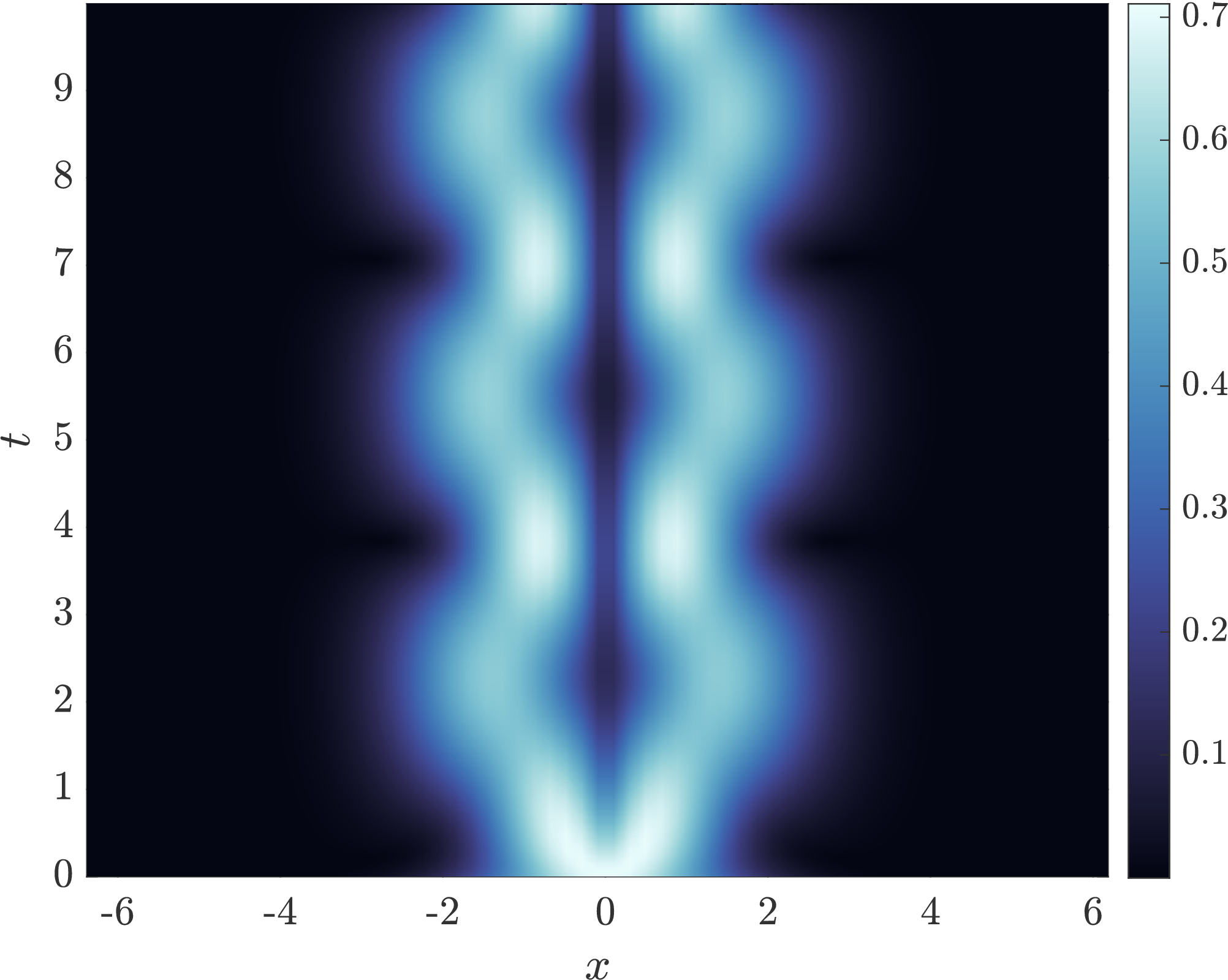}{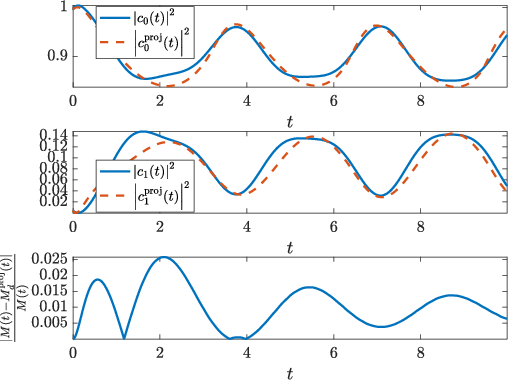}{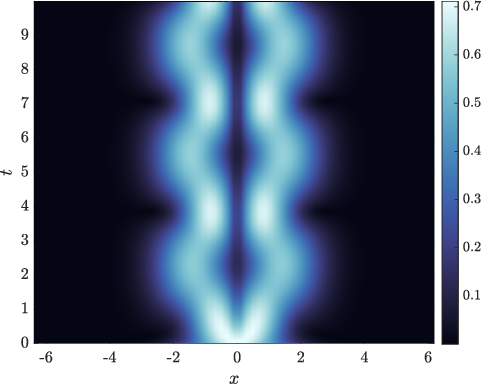}
\caption{The splitting experiment. All conventions here are consistent with Figure~\ref{fig:BECSqueezTest}, except in this case, the dynamics are furnished by the Hamiltonian~\eqref{eq:SplitHam}.}
\label{fig:BECSplitTest}
\end{figure*}

Figures~\ref{fig:BECSqueezTest} and~\ref{fig:BECSplitTest} show the results of these numerical experiments. We present false color plots of $\abs{\psi}^2$, $\abs{\psi_{\rm proj}}^2$, and $\abs{\psi_{\rm galerkin}}^2$. These show excellent qualitative agreement, especially the last two, showing that the main source of disagreement comes from the truncation. They also show strong agreement between $c_j$ and $c_j^{\rm proj}$. In particular, we find good visual agreement in the Rabi frequency, i.e., the peak frequency of mass transfer between the first two even modes. This agreement is exhibited by the similar periodic behavior between $c_n(t),$ determined by System~\eqref{eq:Hamiltons}, and the projected coefficients $c^{\rm proj}_{n}(t)$. Finally, they show a discrepancy of at most $3\%$ between the simulated and projected discrete masses in either experiment. As seen in Figures~\ref{fig:BECSqueezTest} and~\ref{fig:BECSplitTest}, this discrepancy can be mainly attributed to the tails of the distribution $|\psi_{\rm GPE}|^2$.

\begin{figure}[htbp]
\includegraphicsTwoDown{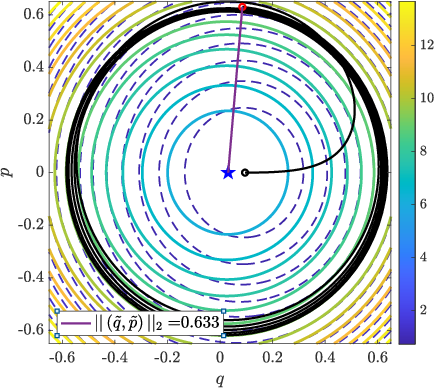}{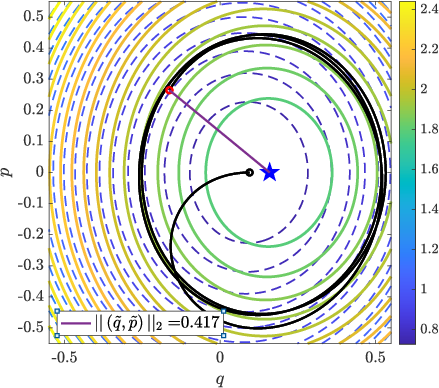}
\caption{Phase portraits for the dynamics due to \textbf{(a)} the squeezing Hamiltonian~\eqref{eq:FinalSqueezeHam}, and \textbf{(b)} the splitting Hamiltonian~\eqref{eq:SplitHamqp}. Dashed contours: the initial Hamiltonian, $\mathcal{H}_0$, initial conditions set at its stable fixed point. Solid contours: the final Hamiltonian, $\mathcal{H}_T$, with a blue star at its minimum. Black lines: numerical trajectories. Red circle: the final state, $(q(T),p(T))$.}
\label{fig:HamContour}
\end{figure}

In visualizing the phase portraits associated with Hamiltonians~\eqref{eq:FinalSqueezeHam} and~\eqref{eq:SplitHam}, we use the same numerical setting as that of Figures~\ref{fig:BECSqueezTest} and~\ref{fig:BECSplitTest}. The phase portraits, shown in Figure~\ref{fig:HamContour}, reveal how significant Rabi oscillations present in Figures~\ref{fig:BECSqueezTest} and~\ref{fig:BECSplitTest} are characterized by the distance between the final state $(q(T),p(T))$ and the stable fixed point $(q^*,p^*)$ of Hamiltonian~\eqref{eq:FinalSqueezeHam}. We denote the initial and final Hamiltonians $\mathcal{H}\left(q(0),p(0),u_0\right)$, $\mathcal{H}\left(q(T),p(T),u_T\right)$ as $\mathcal{H}_0$, $\mathcal{H}_T,$ respectively, in Figure~\ref{fig:HamContour}. 

In conclusion, this section provides numerical justification for the large reduction of dynamic complexity, provided by the truncation of the ansataz~\eqref{eq:GPEGal}. In addition, the reduced dynamics reveal that a successful control strategy should drive the state of the condensate to the global minimum of its finite dimensional Hamiltonian $\mathcal{H}_T$. Furthermore, sub-optimality is almost entirely characterized by the amplitude of simple harmonic motion shown in Figure~\ref{fig:HamContour}.

\section{Optimal Control Framework}
\label{sec:BECoptimization}
We now state an optimal control problem for systems constrained by Hamiltonian dynamics, 
motivated by Sections~\ref{sec:Squeeze} and~\ref{sec:Split}. To this end, we use the admissible class of controls  $\mathcal {U}=\left\{u\in H^1\left([0,T]\right):u(0)=u_0,u(T)=u_T\right\},$ where $u_0,u_T\in\mathbb{R}$ are boundary values for the control $u$. The optimal control problem we study is
\begin{equation}  \label{eq:HamJA}
\min_{u\in \mathcal{U}}{J}=\min_{u\in \mathcal{U}}\left\{\mathcal{H}(q,p,u)\bigg|_{t=T}+\frac{\gamma}{2}\int_0^T\dot{u}^2dt\right\},
\end{equation}
subject to Hamilton's equations
\begin{equation}
\label{eq:HamJAcon}
    \dot{q}=\partial_{p}\mathcal{H},\ q(0)= q_0,\qquad
    \dot{p}=-\partial_{q}\mathcal{H},\ p(0)= p_0.
\end{equation}
Recall, from Section~\ref{sec:intro}, the first term in objective $J,$ called the terminal cost, is used to penalize deviations from some desired state, and that the second term, called the running cost, is a Tikhonov regularization on the control $u$.

Rewriting the terminal cost in~\eqref{eq:HamJA} as a running cost simplifies the process of computing gradients with respect to the state and control variables. We convert terminal costs into running costs through the fundamental theorem of calculus:
\begin{equation*}
\begin{split}
\mathcal{H}\big|_{t=0}^{t=T}&=\int_0^T\frac{d\mathcal{H}}{dt}dt \\
&=\int_0^T\left(\frac{\partial \mathcal{H}}{\partial q}\dot{q}+\frac{\partial \mathcal{H}}{\partial p}\dot{p}+\frac{\partial \mathcal{H}}{\partial u}\dot{u}+\frac{\partial \mathcal{H}}{\partial t}\right)dt  \\
&=\int_0^T\left(\frac{\partial \mathcal{H}}{\partial q}\frac{\partial \mathcal{H}}{\partial p}-\frac{\partial \mathcal{H}}{\partial p}\frac{\partial \mathcal{H}}{\partial q}+\frac{\partial \mathcal{H}}{\partial u}\dot{u}+\frac{\partial \mathcal{H}}{\partial t}\right)dt \\
&=\int_0^T\left(\frac{\partial \mathcal{H}}{\partial u}\dot{u}\right)dt.
\end{split}
\end{equation*}

Using Lagrange multipliers, we express the Hamiltonian optimal control problem in unconstrained form as
\begin{equation}\label{eq:LagrangeBEC}
\begin{split}
\min_{u\in \mathcal{U}}{J} =& 
\min_{u\in \mathcal{U}}
\Biggl\{ \int_0^T
    \Biggl[ \pdv{\mathcal{H}}{u}\dot{u}+\frac{\gamma}{2}\dot{u}^2 + 
    \Biggr.  
\Biggr.
\\
& \phantom{ \min_{u\in \mathcal{U}} \; \Biggl\{ \int_0^T  \Biggl[ \Biggr. \Biggr.}
\Biggl. 
  \Biggr. 
     \lambda^\Tee\left(\dot{q}-\pdv{\mathcal{H}}{p}\right)
        +\mu^\Tee\left(\dot{p}+\pdv{\mathcal{H}}{q}\right)
  \Biggr] \dd t
\Biggr\} \\
=&\min_{u\in \mathcal{U}}
\Biggl\{\int_0^T\mathcal{L}(q,\dot{q},p,\dot{p},u,\dot{u},\lambda,\mu)\dd t
\Biggr\}, 
\end{split}    
\end{equation}
where $^\mathsf{T}$ denotes the matrix transpose, and where the cost $\mathcal{H}\big|_{t=0}$ has been dropped since initial values for the state and control variables are specified, therefore fixed when taking derivatives. 
The necessary conditions for a locally extremal solution to Lagrange problem~\eqref{eq:LagrangeBEC} are given by the Euler-Lagrange equations:
  \begin{align}\label{eq:ConciseCond}
     \frac{d}{dt}&\begin{pmatrix}
           q \\
           p\\
           \lambda \\
           \mu \\
          \gamma\dot{u}
         \end{pmatrix} = 
         A
           \begin{pmatrix}
           \frac{\partial\mathcal{H}}{\partial q} \\
           \frac{\partial\mathcal{H}}{\partial p} \\
           \lambda \\
           \mu \\
          \dot{u}
         \end{pmatrix}-
         \begin{pmatrix}
           0 \\
           0\\
           0 \\
           0 \\
          \frac{d}{dt}\frac{\partial\mathcal{H}}{\partial u}
         \end{pmatrix},\nonumber\\
         &\begin{pmatrix}
           q(0) \\
           p(0)\\
           u(0) 
         \end{pmatrix}=
         \begin{pmatrix}
           q_0 \\
           p_0\\
           u_0 
         \end{pmatrix},
          \begin{pmatrix}
           \lambda(T) \\
           \mu(T)\\
           u(T) 
         \end{pmatrix}=
         \begin{pmatrix}
           0\\
           0\\
           u_T
         \end{pmatrix},
  \end{align}
where 
\begin{align*}
A&=\begin{pmatrix}
    0 & 1 & 0 & 0 & 0 \\
    -1 & 0 & 0 & 0 & 0 \\
    0 & 0 & \frac{\partial^2\mathcal{H}}{\partial q^2} & -\frac{\partial^2\mathcal{H}}{\partial q\partial p} & \frac{\partial^2\mathcal{H}}{\partial q\partial u} \\
    0 & 0 & \frac{\partial^2\mathcal{H}}{\partial p\partial q} & -\frac{\partial^2\mathcal{H}}{\partial p^2} & \frac{\partial^2\mathcal{H}}{\partial p\partial u} \\
    0 & 0 & \frac{\partial^2\mathcal{H}}{\partial u\partial q} & -\frac{\partial^2\mathcal{H}}{\partial u\partial p} & \frac{\partial^2\mathcal{H}}{\partial u^2} \\
\end{pmatrix}\nonumber\\
&=\left(
\begin{array}{c|c}
\mathcal{J} & \mathbf{0}_{2\times3} \\
\hline
 \mathbf{0}_{3\times2} & D\left(\frac{\partial\mathcal{H}}{\partial q}, -\frac{\partial\mathcal{H}}{\partial p}, \frac{\partial\mathcal{H}}{\partial u}\right)
\end{array}
\right),
\end{align*}
with $\mathcal{J}$ denoting the corresponding skew-symmetric matrix and D denoting the Jacobian matrix.

From the perspective of optimal control theory, the equations for $\lambda$ and $\mu$  are called the costate equations and are solved backward in time from their respective terminal conditions. The equation for the control $u(t)$, along with the prescribed boundary conditions, is a two-point boundary value problem. Since solving Equation~\eqref{eq:ConciseCond} in closed form is not possible, we resort to numerical methods discussed in Section~\ref{sec:numerics} in order to solve the optimal control problem~\eqref{eq:HamJA} by approximating the optimality condition~\eqref{eq:ConciseCond}.

\section{Numerical Optimization}\label{sec:numerics}
For the squeezing problem, we use a \emph{hybrid method}: a global, non-convex method followed by a local, iterative method. Hybrid methods, when used appropriately, can overcome non-convexity, yet still remain computationally efficient. 

In both problems, applying the local iterative method requires differentiating the projection coefficients with respect to the control $u$. In the case of the splitting problem, differentiating the coefficients~\eqref{eq:splitprojco} with respect to the control $u$ requires an unmanageable implicit differentiation through Tricomi's function~\eqref{eq:Tricomi} and Equation~\eqref{eq:nuimp}. Therefore, we omit the second optimization step in the splitting problem.

The hybrid method here is similar to work by S{\o}rensen, et al.~\cite{Sorensen}, and allows for the use of a global search routine based on stochastic optimization to overcome non-convexity. 
Global methods are known to converge slowly near a local minimum~\cite{BoydV}. Feeding the result of global methods into a local methods accelerates this slow convergence. 

\subsection{The Global Method}\label{sec:Global}
The first step in the hybrid method reduces the complexity of the optimal control problem so that standard non-convex Nonlinear Programming (NLP) techniques can be applied. This step, the CRAB method~\cite{Doria,Caneva}, constructs the control from the span of an appropriately chosen finite set of basis functions so that the optimization is performed over a small set of unknown coefficients. We choose the basis to ensure that the controls remain in the  admissible space $\mathcal{U}$ of the control problem~\eqref{eq:HamJA}, using
\begin{equation}\label{eq:randclass}
    u_r(t)=\mathcal{P}(t;u_0,u_T,T)+\sum_{j=0}^{N-1}\varepsilon_j\varphi_j(t;T),\quad t\in[0,T],\\
\end{equation}
where $\mathcal{P}$ is a fixed function, $\left\{\varphi_j(t)\right\}_{j\in\mathbb{N}}$ satisfying the boundary conditions defining the admissible class $\mathcal{U}$, is a set of functions that satisfy homogeneous boundary conditions, and the coefficients $\varepsilon_j$ are parameters to be optimized over. 

The CRAB method can be viewed as a Galerkin method, so we must choose the number of basis functions $N$ simultaneously large enough to define an accurate approximation, yet small enough so that the overall procedure remains computationally inexpensive. We have found that a set of 15 basis functions works well.

To solve the resulting NLP problem, we use Differential Evolution (DE)~\cite{Storn}. DE is a stochastic optimization method used to search for candidate solutions to non-convex optimization problems. The idea behind DE is inspired by evolutionary genetics and is thus part of a class of so-called genetic algorithms. 

DE searches the space of candidate solutions by initializing a population set of vectors, known as agents, within some region of the search space. These agents are then mutated (see Algorithm~\ref{algo:mut}) into a new population set, or generation. The mutation operates via two mechanisms: a weighted combination and a random "crossover". 

\begin{figure}[htbp]
\includegraphics[width=.67\textwidth]{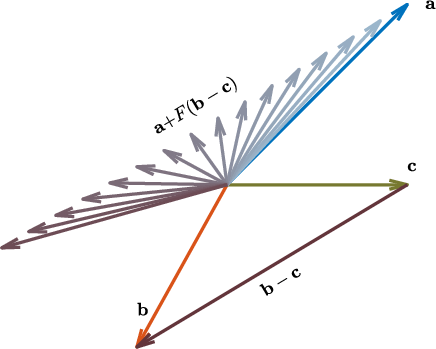}
\caption{Schematic of vectors used to construct the mutation function in Algorithm~\eqref{algo:mut}. The unlabeled vectors are the linear combinations $a+F(b-c),$ for $0.2\le F \le 1.6$, used in the crossover defined in Algorithm~\eqref{algo:mut} and used by Algorithm~\eqref{algo:HDE}.}
\label{fig:Mut}
\end{figure}

At each generation,  Algorithm~\ref{algo:mut}  generates a candidate $z$ to replace each agent $y$. In the mutation step, it chooses at random three agents $a$, $b$, and $c$ to create a new trial agent $\tilde{z}$ through the linear combination
\begin{equation*}
    \tilde{z}=a+F\cdot(b-c),
\end{equation*}
where $F\in[0,2];$ see Figure~\ref{fig:Mut}. In the crossover step, the candidate vector $z$ is constructed by randomly choosing some components  chooses an additional vector $d$ from the current population. 
The new vector $z$ is constructed by randomly choosing some components $\tilde{z}$ and others from an additional randomly-chose agent $d$.
If $J(z)<J(y)$, then $z$ replaces $y$ in the next generation.

DE ensures that the objective functional $J$ of the optimization problem decreases monotonically with (the optimal member of) each generation. As each iteration "evolves" into the next, inferior agents "inherit" optimal traits from superior agents via  mutation, or else are discarded. After a sufficient number of iterations, the best vector in the final generation is chosen as the candidate solution global optimizer.

DE, and genetic algorithms more generally, belong to a class of optimization methods called metaheuristics. Although metaheuristics are useful for non-convex optimization problems, these methods do not guarantee the optimality of candidate solutions. Since the algorithm is stopped after a finite number of iterations, different random realizations return different candidate optimizers. As such, we use DE to search for candidate solutions and use these candidates as initial conditions for a descent method which guarantees local optimality.

\begin{figure}[htbp]
\includegraphicsTwoByTwo{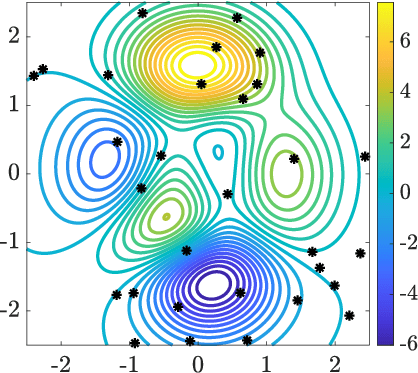}{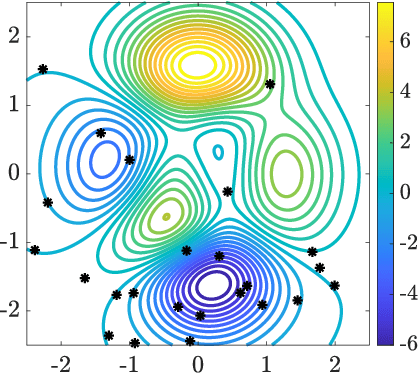}{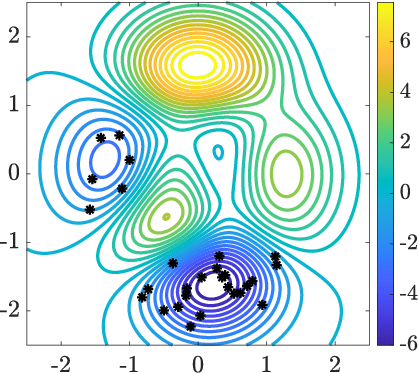}{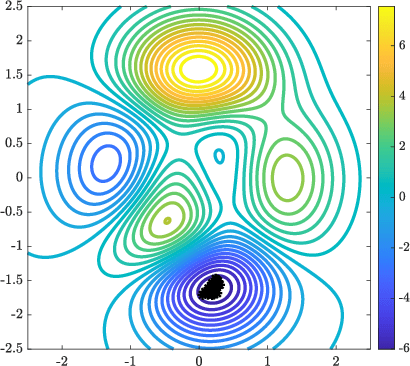}
\caption{Iterates of the DE algorithm applied to the peaks function~\eqref{eq:peaks}:\textbf{(a)}~The initial population. \textbf{(b-d)}~The population after 1, 10, and 20 iterations. Parameters: $F=0.6$ and $R_C=0.9,$; number of agents: $N_{\rm pop} =20$.}
\label{fig:Peaks}
\end{figure}

We show, in Figure~\ref{fig:Peaks}, an example application of DE to minimizing MATLAB's peaks function
\begin{align}\label{eq:peaks}
    f_{\rm peaks}(x,y)&=3(1-x)^2e^{-x^2 - (y+1)^2}-\frac{1}{3}e^{-(x+1)^2-y^2}\nonumber\\&-\left(2x - 10x^3-10y^5\right)e^{-x^2-y^2}
\end{align}
using DE. We see that an initial, random population of vectors converges to the globally optimal regions of the function $f_{\rm peaks}$. At an intermediate generation, the population vectors compete between two local minima, yet the population vectors eventually converge collectively.

We provide a pseudocode of the general method in Algorithm~\ref{algo:HDE}. A more detailed discussion about DE and further implementation and benchmarking details can be found in the book by Storn, et al.~\cite{DiffEvoBook}.

 \begin{algorithm}[htbp]
\caption{Differential Evolution Mutation}\label{algo:mut}
\KwResult{A vector $z$ mutated from agents in a given generation as required by the DE Algorithm~\eqref{algo:HDE}.}
\SetKwInOut{Input}{Input}
 \Input{4 distinct members $a,b,c,d$ from the current generation of agents each with $N$ components, the crossover ratio $R_C\in(0,1),$ and weight $F\in(0,2)$.}
 \For{j=1:N}{
              Compute a random variable $\mathtt{rand}$\;
                 \eIf {$\mathtt{rand}<R_C$} {
                     $z[j]\gets a[j]+F*(b[j]-c[j])$
                     }
                     {$z[j]\gets d[j]$
                 }
         }
\end{algorithm}

 \begin{algorithm}[htbp]
\caption{Differential Evolution}\label{algo:HDE}
\KwResult{A vector likely to be globally optimal with respect to an objective $J$.}
\SetKwInOut{Input}{Input}
 \Input{A maximum number of iterations $\mathtt{Nmax}$, crossover ratio $R_C\in(0,1)$ and weight $F\in(0,2)$}
\While{$\mathtt{counter}<\mathtt{Nmax}$}{
Generate a population \texttt{pop} of $N_{\rm pop}$ vectors.

 \For{$i=1:N_{\rm pop}$}{
 $\mathtt{CurrentMember}\gets \mathtt{Pop}_i$\;
     Choose three distinct vectors $a_i,b_i,c_i$ different from the vector $\mathtt{Pop}_i$\;
        Mutate $a_i,b_i,c_i,$ and the $\mathtt{CurrentMember}$ into the mutated vector $z$
        using the mutation parameters $R_C,F$ and Algorithm~\ref{algo:mut}\;
         \If{$J(z)<J(\mathtt{CurrentMember})$} {
             $\mathtt{TemporaryPop}_i=z$\;
         }
 }
 $\mathtt{Pop}\gets \mathtt{TemporaryPop}$\;
 $\mathtt{counter}\gets \mathtt{counter}+1$\;
}
 \end{algorithm}

\begin{figure}[htbp!]
\includegraphicsTwoDown{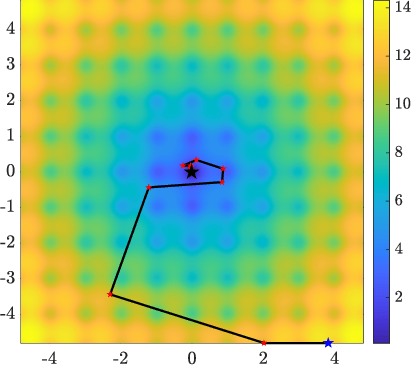}{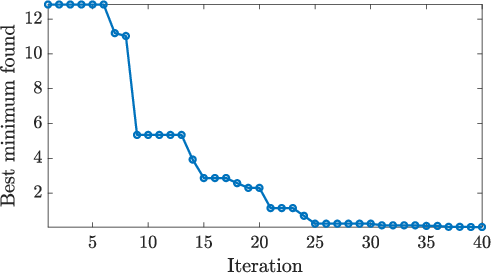}
\caption{Minimization of $f_{\rm Ackley}$~\eqref{eq:Ackley} using the evolutionary Algorithms~\ref{algo:mut} and~\ref{algo:HDE}. \textbf{(a)}~False-color  with the optimal member from each iteration of Algorithm~\ref{algo:HDE} denoted by stars. \textbf{(b)}~The value at the optimal member of each iteration.}
\label{fig:Ack}
\end{figure}

We further demonstrate how DE overcomes non-convexity using a test problem which is much simpler to visualize than the higher dimensional optimal control problem~\eqref{eq:HamJA}. The Ackley function
\begin{align}\label{eq:Ackley}
f_{\rm Ackley}(x,y) = -20&e^{-0.2\sqrt{0.5\left(x^{2}+y^{2}\right)}} \nonumber\\-&e^{0.5\left(\cos 2\pi x + \cos 2\pi y \right)}+e+20,
\end{align}
shown by Figure~\ref{fig:Ack}, is non-convex, w many local minima and a global minimum at the origin. Figure~\ref{fig:Ack} shows the convergence of DE, as outlined in Algorithm~\ref{algo:HDE}, to the global minimum in less than 40 iterations using parameters $N_{\rm pop}=20$, $R_C=0.9$ and $F=0.8$

Using DE with the CRAB method requires  drawing the coefficients $\varepsilon_j$ from the uniform distribution on an appropriately constructed $N$-dimensional hyperrectangle. We choose the half-length of the $j^{\rm th}$ side of the hyperrectangle to decay quadratically as
\begin{equation*}
    l_j=\frac{u_T-u_0}{j^2}.
\end{equation*}
We choose these coefficients to decay quadratically because the Fourier series of an absolutely continuous functions exhibits the same type of decay~\cite{Trefethen}. In this way, the search space of amplitudes $\varepsilon_j$ is not severely restricted, yet the controls generated by the 
CRAB method remain technologically feasible throughout each generation.

\subsection{The Local Method}\label{sec:Local}
For the local search, we use a line search strategy. We introduce here the basic ideas of a line search by discussing the simpler setting of the optimization of a smooth function on $\mathbb{R}^n$, i.e., $ \min_{x\in\mathbb{R}^n}J(x)$. Line searches are iterative methods with two steps per itertation: first, identify a descent direction $p_k,$ and then compute a step size $\alpha_k$ which determines how far $x_k$ should move along $p_k$ at the $k^{\rm th}$ iteration. Put simply, line searches determine $p_k$ and $\alpha_k$ such that
\begin{equation}\label{eq:linesearchfund}
    J(x_{k+1}):=J(x_k+\alpha_kp_k)<J(x_k).
\end{equation}
After a Taylor expansion of Inequality~\eqref{eq:linesearchfund}, we see that
\begin{equation*}
    \left\langle p_k,\nabla J(x_k)\right\rangle_{\mathbb{R}^n}+\mathcal{O}(\alpha_k)<0.
\end{equation*}
For this to hold uniformly in $\alpha_k,$ we should choose the descent direction $p_k$ such that $\left\langle p_k,\nabla J(x_k)\right\rangle_{\mathbb{R}^n}<0$. The most natural choice is $p_k=-\nabla J(x_k),$ in which case the line search is called a $\mathbf{gradient\ descent}$. Choosing $p_k=-H(x_k)^{-1}\nabla J(x_k),$ where $H$ is the Hessian of $J$ and is assumed to be positive-definite, yields a damped Newton-Raphson method.

The task of determining $\alpha_k$ remains. An $\mathit{exact}$ line search chooses $\alpha_k$ to exactly minimize the subproblem
\begin{equation*}
    \min_{\alpha\in\mathbb{R}}J(x_k+\alpha p_k).
\end{equation*}
This is expensive, and it is usually better to allocate resources toward computing better search directions $p_k$ and to approximate the stepsize $\alpha_k$ rather than to determining it exactly. A reasonable approach to choosing $\alpha_k$ is to start with some large value and to then continually reduce it until some criteria is met. Observe that
\begin{equation*}
    J(x_k+\alpha_kp_k)=J(x_k)+\left\langle\alpha_k,p_k\nabla J(x_k)\right\rangle_{\mathbb{R}^n}+\mathcal{O}\left(\alpha_k^2\right).
\end{equation*}
This suggests it is reasonable to decrease $\alpha_k$ until
\begin{equation}\label{eq:agc}
    J(x_k+\alpha_kp_k)\leq J(x_k)+\left\langle\alpha_kp_k\nabla J(x_k)\right\rangle_{\mathbb{R}^n}.
\end{equation}
This $\mathit{inexact}$ line search is called backtracking, and Inequality~\eqref{eq:agc} is called the Armijo-Goldstein condition.

We use the method of gradient descent since Newton's method requires a costly computation of $J's$ second derivatives. Of course, there are many other options to choose from, see e.g.~\cite{BoydV}, but for our purposes, the basic method of gradient descent with Armijo-Goldstein backtracking suffices. 
The last thing we require for the hybrid method is to generalize gradient descent from $\mathbb{R}^n$ to an appropriate affine function space. Von Winckel and Borzi introduced the  \textbf{Gra}dient Descent \textbf{P}ulse \textbf{E}ngineering (GRAPE) algorithm~\cite{vonWinckel}, which  automatically preserves the boundary conditions of the admissible class $\mathcal{U}$ for the optimal control problem~\eqref{eq:HamJA}. It has been used frequently in the quantum control literature; see, e.g.,~\cite{Hohenester,Mennemann,Nature,BorziBook}. The update in the GRAPE method is
\begin{equation}\label{eq:graddesc}
u_{k+1}=u_k-\alpha_k\nabla_{u}\mathcal{L}
\big|_{u=u_k},
\end{equation} 
where the stepsize $\alpha$ is chosen using backtracking, and the Armijo-Goldstein condition for this problem reads
\begin{equation}\label{eq:agbls}
J\left[u_k-\alpha\nabla_{u_k} \mathcal{L}\left(u_k\right)\right]<J[u_k]-\frac{\alpha}{2}\left\|\nabla_{u_k}\mathcal{L}(u_k)\right\|_{L^2([0,T])}^2.
\end{equation}
Until condition~\eqref{eq:agbls} is satisfied, the value of the stepsize $\alpha$ is decreased by some  factor $\phi<1$. Since the gradient descent~\eqref{eq:graddesc} depends on the function space in which $\nabla_{u}\mathcal{L}(u)$ is to be understood, we review some basic facts about calculus on infinite-dimensional (affine) spaces.

The Gateaux derivative of a functional $J$, evaluated at a point 
$u\in\mathcal{U}$ in the direction of a displacement vector $v\in C_c^{\infty}([0,T])$ is defined by
\begin{equation*}
d_{u}J[u;v]:=\lim_{\varepsilon\to0}\frac{J[u+\varepsilon v]-J[u]}{\varepsilon},
\end{equation*}
and if this exists for all admissible displacement vectors $v$, the functional $J$ is said to be Gateaux differentiable. Given the uniform bound $\sup_{u\in\mathcal{U}}\abs{\mathcal{L}(u)}\leq M$ for some finite $M,$ a direct calculation shows
\begin{equation*}
\begin{split}
d_{u}J[u;v]&=\lim_{\varepsilon\to0}\frac{J[u+\varepsilon v]-J[u]}{\varepsilon}  \\
&=\lim_{\varepsilon\to0}\frac{1}{\varepsilon}\left(\int_0^T\mathcal{L}(u+\varepsilon v)dt-\int_0^T\mathcal{L}(u)dt\right) \\
&=\lim_{\varepsilon\to0}\frac{1}{\varepsilon}\int_0^T\int_0^1d_s\mathcal{L}(u+s\varepsilon v)dsdt \\
&=\lim_{\varepsilon\to0}\int_0^T\int_0^1\mathcal{L}'(u+s\varepsilon v)v ds dt \\
&=\int_0^T\nabla_{u}\mathcal{L}(u)v dt:=\left\langle\delta_uJ,v\right\rangle_{L^2([0,T])},
\end{split}
\end{equation*}
using the bound on $\mathcal{L}$ in order to invoke the Lebesgue dominated convergence theorem in the last equality.

The gradient of $\mathcal{L}$ with respect to the $L^2([0,T])$ inner product can be identified with the functional derivative $\delta_uJ$ calculated and expressed through the last entry in Equation \eqref{eq:ConciseCond}, i.e., $ \nabla_u\mathcal{L}=\delta_uJ$ in the space $L^2([0,T])$. However, were one to perform a gradient descent on an initially admissible control $u_k$, the increment $\alpha_k\nabla_{u}\mathcal{L}
\big|_{u=u_k}$ would fail to satisfy the boundary conditions and the updated function would leave the admissible set $\mathcal{U}$. We can avoid this problem by drawing the update from a more carefully chosen function space. 

Since Taylor's theorem must hold for all sufficiently regular functionals on Hilbert spaces, the Taylor series
\begin{align*}
J[u+\varepsilon v]&=J[u]+\varepsilon d_{u}J[u,v]+\mathcal{O}(\varepsilon^2)\nonumber\\&=J[u]+\varepsilon\left\langle\nabla_{u}\mathcal{L}(u),v\right\rangle_{X}+\mathcal{O}(\varepsilon^2)
\end{align*}
holds term-by-term for all spaces $X$. The vWB method relies on choosing $X$ to be the traceless and homogeneous Sobolev space $\dot{H}_0^1([0,T]),$ i.e., the vector space of measurable functions, that vanish on the boundary of [0,T] such that the norm $\abs{\abs{\star}}_{\dot{H}^1([0,T])}$ induced by the inner product
\begin{equation*}
\left\langle\star,\star\right\rangle_{\dot{H}^1([0,T])}:=\int_0^T\left(\partial_t{\star}\right)^{\dag}\left(\partial_t{\star}\right)dt
\end{equation*}
is finite.
This implies, by equating the Gateaux differential with respect to $L^2([0,T])$ and with respect to $\dot{H}_0^1([0,T])$,
\begin{align}\label{eq:weakform}
\left\langle\delta_u J,v\right\rangle_{L^2([0,T])}&=\left\langle\nabla_{u}\mathcal{L}(u),v\right\rangle_{\dot{H}_0^1([0,T])}\nonumber\\&
=\int_0^T \partial_t\nabla_{u}\mathcal{L}\partial_tvdt\nonumber\\&
=-\int_0^T\partial_t^2\nabla_{u}\mathcal{L}v dt\nonumber\\&=-\left\langle\partial_t^2\nabla_{u}\mathcal{L},v \right\rangle_{L^2([0,T])},
\end{align}
where integration by parts is used once along with the boundary conditions of $v$. Since this holds for all displacements $v\in C_c^{\infty}([0,T])$, we conclude that in order to perform a gradient descent at the current control $u$, we must first solve the strong form of~\eqref{eq:weakform}
\begin{equation} \label{eq:Poisson}
\partial_t^2\nabla_{u}\mathcal{L} =-\delta_u J,\  \ \nabla_u \mathcal{L}(0)=\nabla_u \mathcal{L}(T)=0
\end{equation}
in order to determine the admissible gradient of the objective with respect to the control. 

Note that boundary value problem~\eqref{eq:Poisson} yields a control gradient with homogeneous Dirichlet boundary conditions. This implies that the use of an iterative method which uses this control gradient in an update automatically preserves the boundary conditions of the control, as desired. In order to solve the two-point boundary values problems for the control gradients, we use spectral methods such as Chebyshev collocation~\cite{Trefethen}.

We provide a straightforward extension to the GRAPE method, appropriate for problems where, in addition, Neumann boundary data is specified for the admissible class $\mathcal{U}$. We encounter a problem of this type in subsection~\ref{sec:Squeeze}. The idea is to use the inner product on $\dot{H}_0^2([0,T])$, so that  we are instead tasked with solving an inhomogeneous biharmonic equation with homogeneous boundary data:
\begin{align*} 
&\partial_t^4\nabla_{u}\mathcal{L} =\delta_u J,\nonumber\\&  \ \nabla_u \mathcal{L}(0)=\nabla_u \mathcal{L}(T)=\partial_t\nabla_{u}\mathcal{L}\big|_{t=0}=\partial_t\nabla_{u}\mathcal{L}\big|_{t=T}=0.
\end{align*}
Once again, the gradient $\nabla_{u}\mathcal{L}(u)$ preserves the appropriate boundary data when using a line search. In fact, it is also clear that the boundary value problem
\begin{align} \label{eq:pharmonic}
&\partial_t^{2p}\nabla_{u}\mathcal{L}=(-1)^p\delta_u J,\nonumber\\&  \partial_t^j\nabla_{u}\mathcal{L}\big|_{t=0}=\partial_t^j\nabla_{u}\mathcal{L}\big|_{t=T}=0,\ j=0,1,\ldots,p-1,
\end{align}
generalizes the GRAPE method to the space $\dot{H}_0^p([0,T]),$ for $p\in\mathbb{Z}^+$. This method is summarized by Algorithm~\eqref{algo:BPGD}.

\begin{algorithm}
\caption{Gradient Descent Method in $\dot{H}_0^p([0,T])$.}\label{algo:BPGD}
\KwResult{Admissible control $u$ which is locally optimal with respect to the objective functional}
\SetKwInOut{Input}{Input}
 \Input{Initial admissible control $u$, the objective functional $J$, tolerance $\mathtt{tol}$, maximum number of iterations $\mathtt{Nmax}$, and reduction parameter $r\in(0,1)$}
\While{$ \mathtt{error}> \mathtt{tol}$ \rm and $\mathtt{counter}< \mathtt{Nmax}$}{
Evolve the state variable $(q,p)$ from $t=0$ to $t=T$, using Equations~\eqref{eq:HamJAcon}\;
Evolve the costate variables $(\lambda,\mu)$ from $t=T$ to $t=0$, using Equation~\eqref{eq:ConciseCond}\;
Compute $\nabla_{u}\mathcal{L}$ via Equation~\eqref{eq:pharmonic} with source term $\frac{d}{dt}\frac{\partial\mathcal{H}}{\partial u}$ given by Equation~\eqref{eq:ConciseCond}\;

 \While {\rm Inequality~\eqref{eq:agbls} is false and $\alpha>\mathtt{tol}$}{
 $\alpha\gets r\alpha$\;
}
\eIf{$\alpha<\mathtt{tol}$}{
\texttt{break};}
{
$u\gets u-\alpha\nabla_{u}\mathcal{L}$\;
$\mathtt{error}\gets J[u]-J[u+\alpha\nabla_{u}\mathcal{L}]$\;
$\mathtt{counter}\gets \mathtt{counter}+1$\;
}
}
\end{algorithm}

\section{Numerical Results}\label{sec:Results}

We use the optimization methods of Section~\ref{sec:BECoptimization} in order to numerically solve the optimal control problem~\eqref{eq:HamJA}. We find that using the Tikhonov parameter $\gamma=10^{-4}$ is sufficient to render the problem well-conditioned. The global CRAB/DE method, being a search method, provides no guarantee of reaching the global minimizer. Indeed, we have found several results that are competitive in minimizing the objective. 

\section*{The squeezing problem} 
To demonstrate the necessity of the hybrid method, we first demonstrate what happens when we omit the first step, the global search. We apply the descent method, Algorithm~\eqref{algo:BPGD} with $p=1,$ directly to the squeezing problem from Section~\ref{sec:Squeeze} in order to compute a local minimum downhill from the linear ramp used in Figure~\ref{fig:BECSqueezTest}.

\begin{figure*}[htbp]
\includegraphicsTwoByTwo{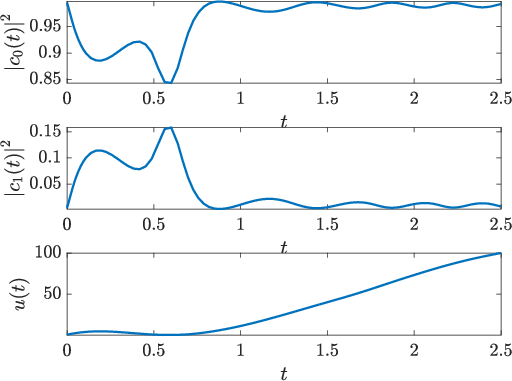}{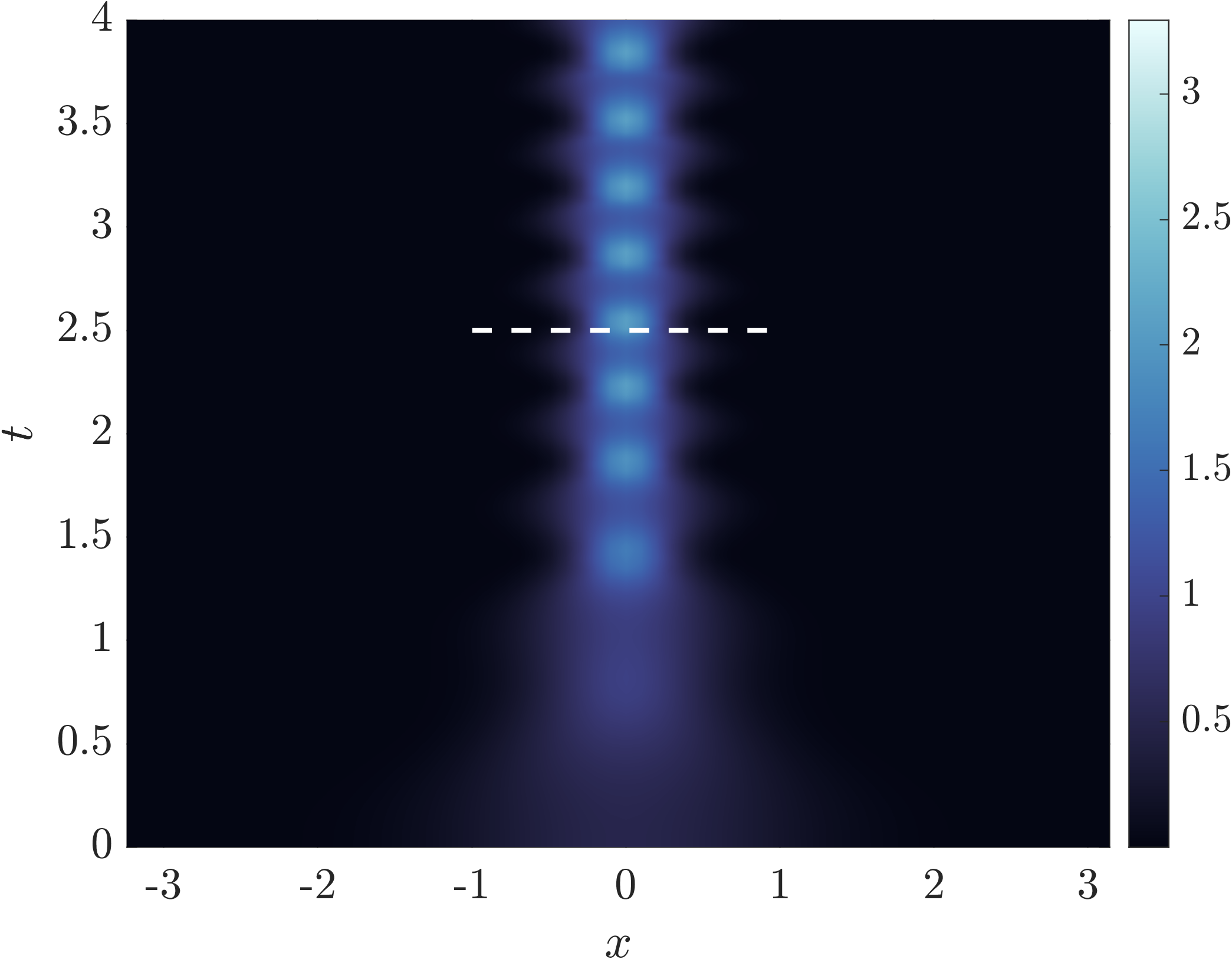}{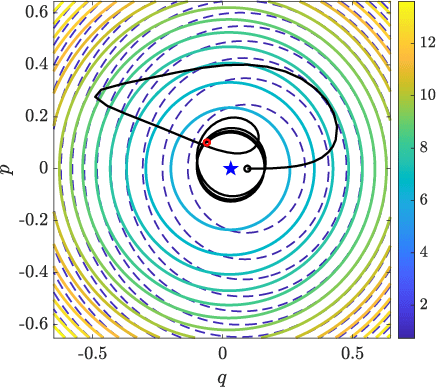}{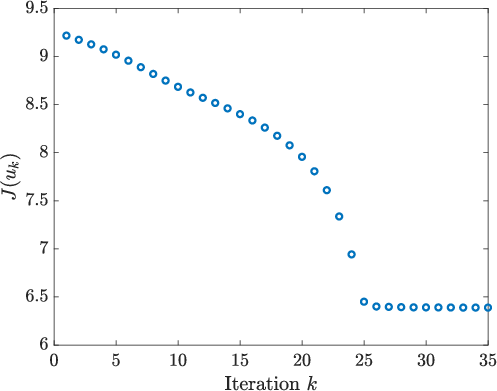}
\caption{The result of using the GRAPE algorithm of Section~\ref{sec:numerics} in the space $\dot{H}_0^1([0,T])$ with the linear ramp from Figure~\ref{fig:BECSqueezTest} as an initial control. Conventions used here are similar to conventions used in Figures~\ref{fig:BECSqueezTest} and~\ref{fig:HamContour}. \textbf{(a)}~The Galerkin coefficients which satisfy Equations~\eqref{eq:GPEGal} with optimal control $u$. \textbf{(b)}~The numerical solution of the GPE with the optimal control $u$ from Panel (a) up until the dashed white line. The persisting dynamics are computed with constant control $u(T)$. \textbf{(c)}~The resulting phase portrait with the inset showing the persistent oscillation.  \textbf{(d)}~The convergence of GRAPE.}
\label{fig:BECIndSqueezResult}
\end{figure*}

Figure~\ref{fig:BECIndSqueezResult}, shows the locally optimal control, the locally optimal state dynamics, the corresponding numerical solution of the GPE~\eqref{eq:GPE}, the corresponding phase portrait for the reduced dynamics, and convergence of GRAPE. The procedure is moderately successful, approximately halving the objective function, but it fails to eliminate a significant oscillation.

We now show the results of the full hybrid method. 
For the CRAB ansatz, we use 15 sine modes and an admissible linear ramp, setting
\begin{equation}\label{eq:pracCRAB}
    u_r(t)=u_0+(u_T-u_0)\frac{t}{T}+\sum_{j=1}^{15}\varepsilon_j\sin\left(\frac{j\pi t}{T}\right)
\end{equation}
We apply DE to determine effective coefficients $\varepsilon_j$, with parameters  $F=0.8,\ R_C=0.9,\ N_P=40,$ and $N_{\rm max}=30$ in Algorithms~\eqref{algo:mut} and~\eqref{algo:HDE}. 

Note that the value of $\mathcal{H}_T$ depends on the quantity $\dot{u}\big|_{t=T}$. Since we are interested in the case that $\mathcal{H}$ is constant for $t>T$, the minimum value of the Hamiltonian we are truly interested in is independent of any terms which depend on the derivative of the control. For this reason, we choose to minimize the Hamiltonian with $\dot{u}$ set to 0 at $t=T$. We follow this with a GRAPE descent algorithm in $\dot{H}_0^2([0,T]),$ i.e. Algorithm~\eqref{algo:BPGD} with $p=2$ in order to preserve both Dirchlet \textit{and} Neumann data. This allows us to perform a line search for controls that minimize the modified Hamiltonian ${\left.\mathcal{H}_T\right|}_{\dot{u}=0},$ rather than the full terminal Hamiltonian $\mathcal{H}_T$.

\begin{figure*}[htbp]
\includegraphicsTwoByTwo{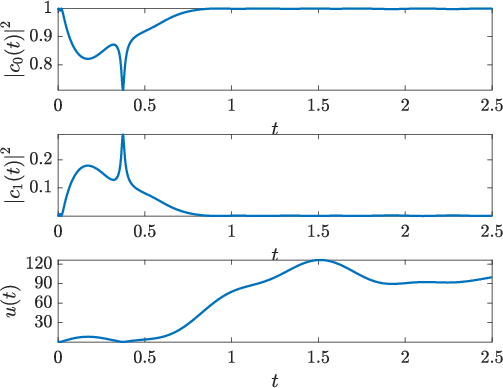}{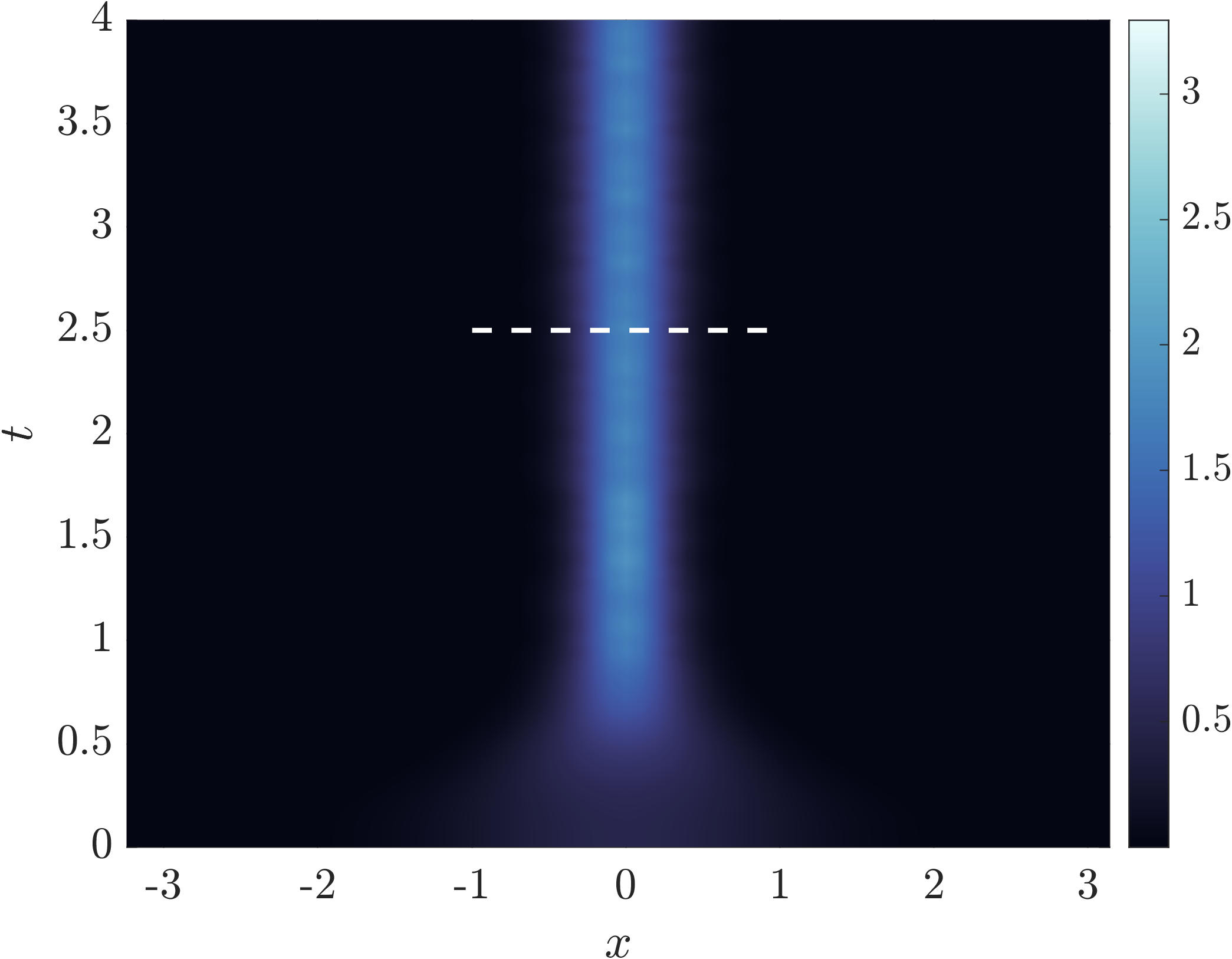}{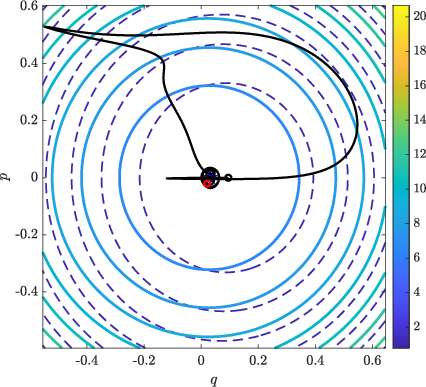}{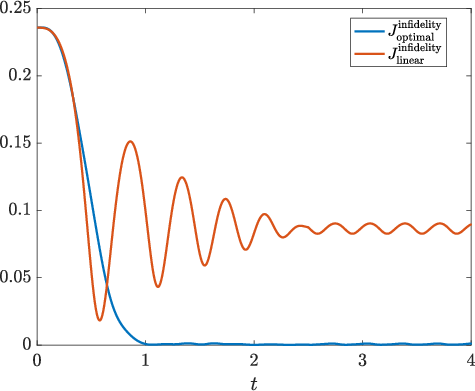}
\caption{The result of using the hybrid optimization technique outlined in Section~\ref{sec:numerics} on the squeezing problem of Subsection~\ref{sec:Squeeze}. The conventions used here are identical to those used in Figure~\ref{fig:BECIndSqueezResult}. Panel (d) shows the infidelity~\eqref{eq:Jdef} of the optimal control and infidelity of the linear control from Figure~\ref{fig:BECSqueezTest}.}
\label{fig:BECSqueezResult}
\end{figure*}

The hybrid method performs significantly better, as seen in Figure~\ref{fig:BECSqueezResult}. The value of the terminal Hamiltonian ${\left.\mathcal{H}_T\right|}_{\dot{u}=0}$ is one order of magnitude smaller than the terminal Hamiltonian in Figure~\ref{fig:BECIndSqueezResult}. To further characterize optimality, we compute the infidelity term from the Hohenester objective~\eqref{eq:Jdef} between the computed solution of the GPE  and the at time $t=T$, and $\psi_d$ the minimizer of the GPE energy.
Figure~\ref{fig:BECSqueezResult} shows that the hybrid method has reduced $J^{\rm infidelity}$  by an order of magnitude compared the linearly controlled condensate. 

\begin{figure*}[htbp]
\includegraphicsTwoByTwo{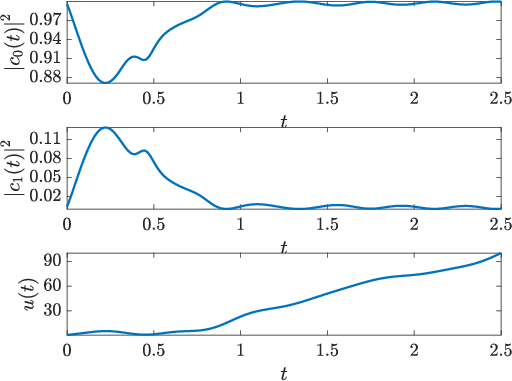}{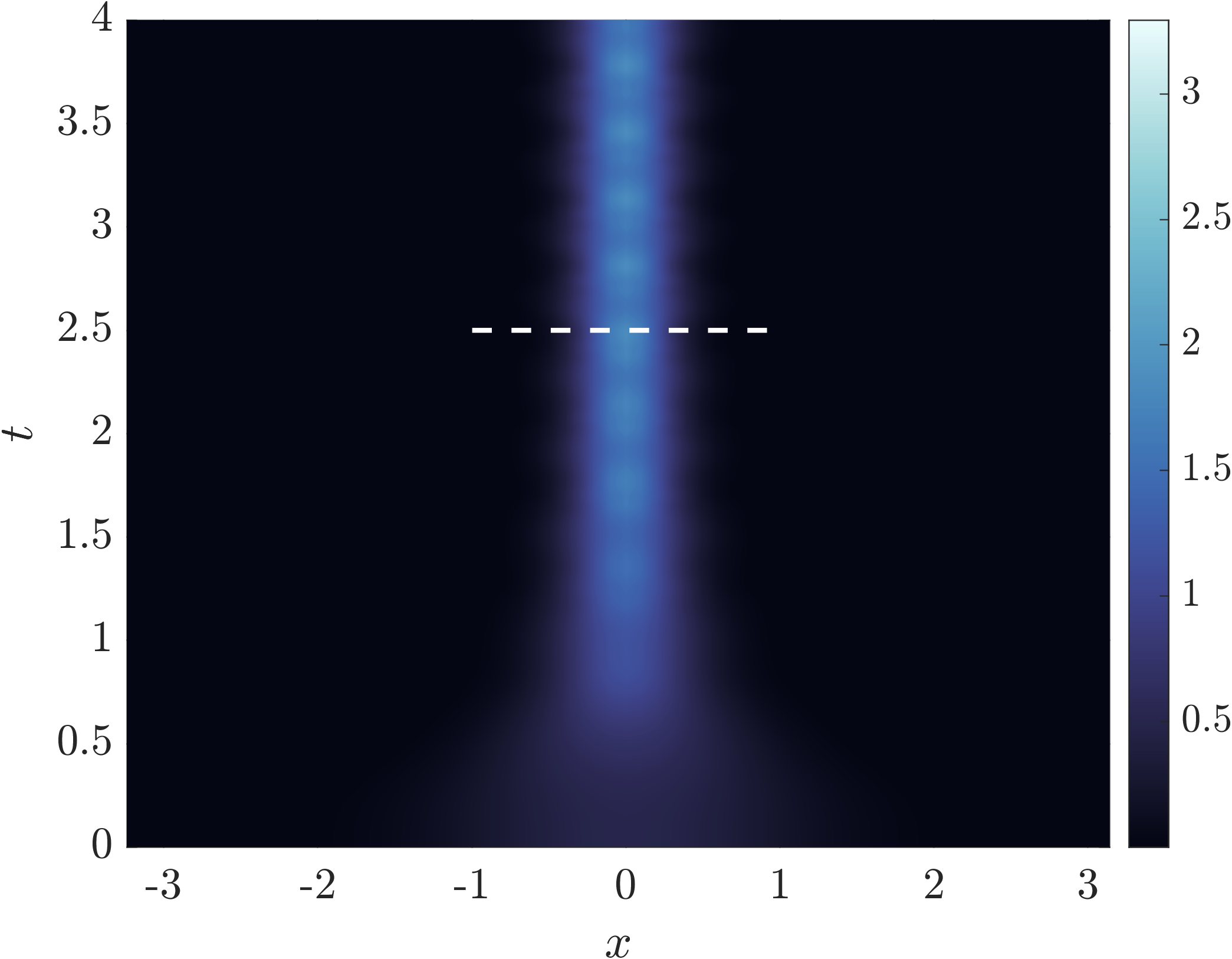}{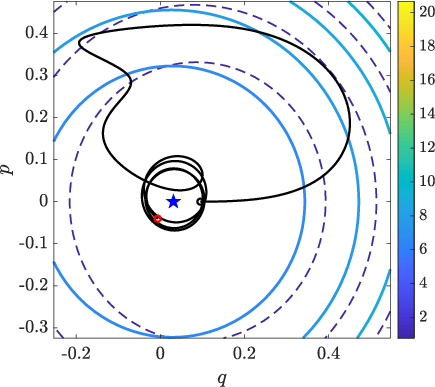}{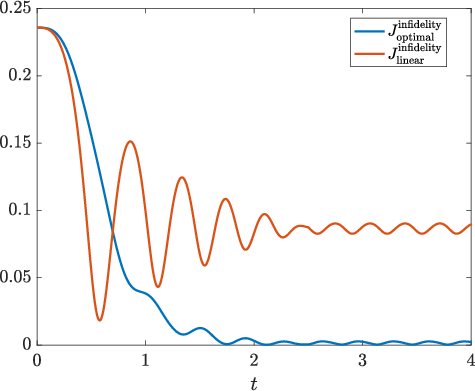}
\caption{Another result of using the hybrid optimization technique outlined in Section~\ref{sec:numerics} on the squeezing problem of Subsection~\ref{sec:Squeeze}. The conventions used here are identical to those used in Figure~\ref{fig:BECIndSqueezResult}. Panel (d) shows the infidelity~\eqref{eq:Jdef} of the optimal control and infidelity of the linear control from Figure~\ref{fig:BECSqueezTest}.}
\label{fig:BECSqueezResult2}
\end{figure*}

We notice the coefficients $c_0$ and $c_1$ resulting from the hybrid method, shown in Figure~\ref{fig:BECSqueezResult}, lose a fair amount of regularity at certain moments during the control process. For this reason, we show another, slightly less optimal, result in Figure~\ref{fig:BECSqueezResult2} found by the same methodology, but where the dynamics are smoother. Note that the dynamics lose smoothness at the precise instant that $u$ becomes very small. In a technological setting, these irregularities can be more systematically avoided by appending an inequality constraint to the admissible space $\mathcal{U},$ or by further using a Tikhonov regularization on the dynamics in the objective~\eqref{eq:HamJA}.

\section*{The splitting problem}
\begin{figure*}[htbp]
\includegraphicsTwoByTwo{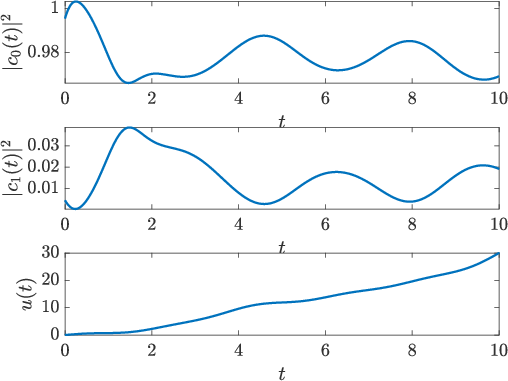}{figures/bestSplit3.png}{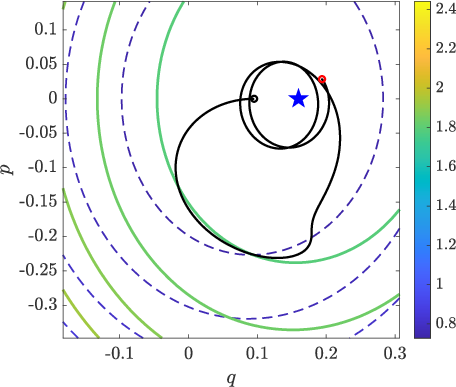}{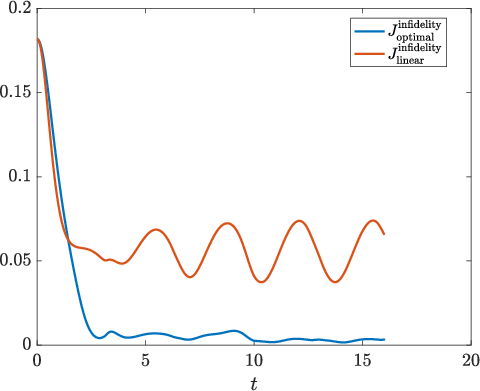}
\caption{The result of using the CRAB method on the splitting problem of Section~\ref{sec:Squeeze}. The same conventions of Figures~\ref{fig:BECSplitTest} and~\ref{fig:BECSqueezResult} are used here.}
\label{fig:BECSplitResult}
\end{figure*}

Figure~\ref{fig:BECSplitResult} shows similar results for the splitting problem of Section~\ref{sec:Split}. All conventions used here are the same as those of Figure~\ref{fig:BECSplitTest}. As discussed in Section~\ref{sec:numerics}, we perform only the global optimization and not the descent method.
Despite not applying a descent method, the global method significantly reduces the oscillations compared with the linear ramp control.

 The squeezing problem takes an average of about thirty second on a 2.6 GHz 6-Core Intel i7 Macbook Pro, while the splitting problem takes 3-5 minutes. The splitting problem takes more time since for each evaluation of the objective in Equation~\eqref{eq:HamJA} it must compute the costly inner products of Equations~\eqref{eq:splitprojco}.

\section{Conclusion}

We have demonstrated that reducing the GPE dynamics to a single non-autonomous degree of freedom Hamiltonian system and assuming a restricted form of the control is an effective and inexpensive approach to optimal control of two problems in the reshaping of a Bose-Einstein condensate. Moreover, we provide a complete characterization of the physics of controlled condensates using standard dynamical systems techniques. The techniques described here can be applied to other control problems constrained by Hamiltonian PDE, and perhaps to problems where posing an optimal control problem is challenging, if not impossible, without a visualization in a low dimensional setting.

Further refinements of this work can be pursued. This includes generalizing the form of the potentials shown in the GPE~\eqref{eq:GPE} so that, for example, the optimization is performed both over space and time. Also, a truncation of the Galerkin expansion~\eqref{eq:GPEGal} at a higher order can be pursued to refine the suppression of excitations that might have been missed by the order of the reduction used in this paper. While we have applied the Galerkin truncation to the GPE in one space dimension, the approach should still be applicable to problems in two or three dimensions. The speedup enabled by this reduction allows the use of DE, which, in turn, permits the exploration of a large class of controls.

\section*{Acknowledgments}

We would like to thank Professors David Shirokoff, Christina Frederick, Alejandro Aceves, Richard Moore, and Panos Kevrekedis for their many helpful comments and discussions.

\bibliographystyle{apsrev4-1} 
\bibliography{bibliography.bib} 

\begin{thebibliography}{34}%
\makeatletter
\providecommand \@ifxundefined [1]{%
 \@ifx{#1\undefined}
}%
\providecommand \@ifnum [1]{%
 \ifnum #1\expandafter \@firstoftwo
 \else \expandafter \@secondoftwo
 \fi
}%
\providecommand \@ifx [1]{%
 \ifx #1\expandafter \@firstoftwo
 \else \expandafter \@secondoftwo
 \fi
}%
\providecommand \natexlab [1]{#1}%
\providecommand \enquote  [1]{``#1''}%
\providecommand \bibnamefont  [1]{#1}%
\providecommand \bibfnamefont [1]{#1}%
\providecommand \citenamefont [1]{#1}%
\providecommand \href@noop [0]{\@secondoftwo}%
\providecommand \href [0]{\begingroup \@sanitize@url \@href}%
\providecommand \@href[1]{\@@startlink{#1}\@@href}%
\providecommand \@@href[1]{\endgroup#1\@@endlink}%
\providecommand \@sanitize@url [0]{\catcode `\\12\catcode `\$12\catcode
  `\&12\catcode `\#12\catcode `\^12\catcode `\_12\catcode `\%12\relax}%
\providecommand \@@startlink[1]{}%
\providecommand \@@endlink[0]{}%
\providecommand \url  [0]{\begingroup\@sanitize@url \@url }%
\providecommand \@url [1]{\endgroup\@href {#1}{\urlprefix }}%
\providecommand \urlprefix  [0]{URL }%
\providecommand \Eprint [0]{\href }%
\providecommand \doibase [0]{http://dx.doi.org/}%
\providecommand \selectlanguage [0]{\@gobble}%
\providecommand \bibinfo  [0]{\@secondoftwo}%
\providecommand \bibfield  [0]{\@secondoftwo}%
\providecommand \translation [1]{[#1]}%
\providecommand \BibitemOpen [0]{}%
\providecommand \bibitemStop [0]{}%
\providecommand \bibitemNoStop [0]{.\EOS\space}%
\providecommand \EOS [0]{\spacefactor3000\relax}%
\providecommand \BibitemShut  [1]{\csname bibitem#1\endcsname}%
\let\auto@bib@innerbib\@empty
\bibitem [{\citenamefont {Glaser}\ \emph {et~al.}(2015)\citenamefont {Glaser},
  \citenamefont {Boscain}, \citenamefont {Calarco}, \citenamefont {Koch},
  \citenamefont {K{\"o}ckenberger}, \citenamefont {Kosloff}, \citenamefont
  {Kuprov}, \citenamefont {Luy}, \citenamefont {Schirmer}, \citenamefont
  {Schulte-Herbr{\"u}ggen} \emph {et~al.}}]{glaser2015training}%
  \BibitemOpen
  \bibfield  {author} {\bibinfo {author} {\bibfnamefont {S.~J.}\ \bibnamefont
  {Glaser}}, \bibinfo {author} {\bibfnamefont {U.}~\bibnamefont {Boscain}},
  \bibinfo {author} {\bibfnamefont {T.}~\bibnamefont {Calarco}}, \bibinfo
  {author} {\bibfnamefont {C.~P.}\ \bibnamefont {Koch}}, \bibinfo {author}
  {\bibfnamefont {W.}~\bibnamefont {K{\"o}ckenberger}}, \bibinfo {author}
  {\bibfnamefont {R.}~\bibnamefont {Kosloff}}, \bibinfo {author} {\bibfnamefont
  {I.}~\bibnamefont {Kuprov}}, \bibinfo {author} {\bibfnamefont
  {B.}~\bibnamefont {Luy}}, \bibinfo {author} {\bibfnamefont {S.}~\bibnamefont
  {Schirmer}}, \bibinfo {author} {\bibfnamefont {T.}~\bibnamefont
  {Schulte-Herbr{\"u}ggen}},  \emph {et~al.},\ }\href@noop {} {\bibfield
  {journal} {\bibinfo  {journal} {Eur. Phys. J. D}\ }\textbf {\bibinfo {volume}
  {69}},\ \bibinfo {pages} {1} (\bibinfo {year} {2015})}\BibitemShut {NoStop}%
\bibitem [{\citenamefont {Borzi}\ \emph {et~al.}(2017)\citenamefont {Borzi},
  \citenamefont {Ciaramella},\ and\ \citenamefont {Sprengel}}]{BorziBook}%
  \BibitemOpen
  \bibfield  {author} {\bibinfo {author} {\bibfnamefont {A.}~\bibnamefont
  {Borzi}}, \bibinfo {author} {\bibfnamefont {G.}~\bibnamefont {Ciaramella}}, \
  and\ \bibinfo {author} {\bibfnamefont {M.}~\bibnamefont {Sprengel}},\
  }\href@noop {} {\emph {\bibinfo {title} {Formulation and Numerical Solution
  of Quantum Control Problems}}}\ (\bibinfo  {publisher} {SIAM},\ \bibinfo
  {address} {Philadeplphia},\ \bibinfo {year} {2017})\BibitemShut {NoStop}%
\bibitem [{\citenamefont {Anderson}\ \emph {et~al.}(1995)\citenamefont
  {Anderson}, \citenamefont {Ensher}, \citenamefont {Matthews}, \citenamefont
  {Wieman},\ and\ \citenamefont {Cornell}}]{Weiman}%
  \BibitemOpen
  \bibfield  {author} {\bibinfo {author} {\bibfnamefont {M.~H.}\ \bibnamefont
  {Anderson}}, \bibinfo {author} {\bibfnamefont {J.~R.}\ \bibnamefont
  {Ensher}}, \bibinfo {author} {\bibfnamefont {M.~R.}\ \bibnamefont
  {Matthews}}, \bibinfo {author} {\bibfnamefont {C.~E.}\ \bibnamefont
  {Wieman}}, \ and\ \bibinfo {author} {\bibfnamefont {E.~A.}\ \bibnamefont
  {Cornell}},\ }\href {\doibase 10.1126/science.269.5221.198} {\bibfield
  {journal} {\bibinfo  {journal} {Science}\ }\textbf {\bibinfo {volume}
  {269}},\ \bibinfo {pages} {198} (\bibinfo {year} {1995})}\BibitemShut
  {NoStop}%
\bibitem [{\citenamefont {Bradley}\ \emph {et~al.}(1995)\citenamefont
  {Bradley}, \citenamefont {Sackett}, \citenamefont {Tollett},\ and\
  \citenamefont {Hulet}}]{PhysRevLett.75.1687}%
  \BibitemOpen
  \bibfield  {author} {\bibinfo {author} {\bibfnamefont {C.~C.}\ \bibnamefont
  {Bradley}}, \bibinfo {author} {\bibfnamefont {C.~A.}\ \bibnamefont
  {Sackett}}, \bibinfo {author} {\bibfnamefont {J.~J.}\ \bibnamefont
  {Tollett}}, \ and\ \bibinfo {author} {\bibfnamefont {R.~G.}\ \bibnamefont
  {Hulet}},\ }\href {\doibase 10.1103/PhysRevLett.75.1687} {\bibfield
  {journal} {\bibinfo  {journal} {Phys. Rev. Lett.}\ }\textbf {\bibinfo
  {volume} {75}},\ \bibinfo {pages} {1687} (\bibinfo {year}
  {1995})}\BibitemShut {NoStop}%
\bibitem [{\citenamefont {Davis}\ \emph {et~al.}(1995)\citenamefont {Davis},
  \citenamefont {Mewes}, \citenamefont {Andrews}, \citenamefont {van Druten},
  \citenamefont {Durfee}, \citenamefont {Kurn},\ and\ \citenamefont
  {Ketterle}}]{PhysRevLett.75.3969}%
  \BibitemOpen
  \bibfield  {author} {\bibinfo {author} {\bibfnamefont {K.~B.}\ \bibnamefont
  {Davis}}, \bibinfo {author} {\bibfnamefont {M.~O.}\ \bibnamefont {Mewes}},
  \bibinfo {author} {\bibfnamefont {M.~R.}\ \bibnamefont {Andrews}}, \bibinfo
  {author} {\bibfnamefont {N.~J.}\ \bibnamefont {van Druten}}, \bibinfo
  {author} {\bibfnamefont {D.~S.}\ \bibnamefont {Durfee}}, \bibinfo {author}
  {\bibfnamefont {D.~M.}\ \bibnamefont {Kurn}}, \ and\ \bibinfo {author}
  {\bibfnamefont {W.}~\bibnamefont {Ketterle}},\ }\href {\doibase
  10.1103/PhysRevLett.75.3969} {\bibfield  {journal} {\bibinfo  {journal}
  {Phys. Rev. Lett.}\ }\textbf {\bibinfo {volume} {75}},\ \bibinfo {pages}
  {3969} (\bibinfo {year} {1995})}\BibitemShut {NoStop}%
\bibitem [{\citenamefont {Pitaevskii}\ and\ \citenamefont
  {Stringari}(2003)}]{GPEref}%
  \BibitemOpen
  \bibfield  {author} {\bibinfo {author} {\bibfnamefont {L.~P.}\ \bibnamefont
  {Pitaevskii}}\ and\ \bibinfo {author} {\bibfnamefont {S.}~\bibnamefont
  {Stringari}},\ }\href@noop {} {\emph {\bibinfo {title} {Bose-{E}instein
  Condensation}}}\ (\bibinfo  {publisher} {Clarendon Press},\ \bibinfo
  {address} {Oxford},\ \bibinfo {year} {2003})\BibitemShut {NoStop}%
\bibitem [{\citenamefont {Gross}\ \emph {et~al.}(2010)\citenamefont {Gross},
  \citenamefont {Zibold}, \citenamefont {Nicklas}, \citenamefont {Esteve},\
  and\ \citenamefont {Oberthaler}}]{gross2010}%
  \BibitemOpen
  \bibfield  {author} {\bibinfo {author} {\bibfnamefont {C.}~\bibnamefont
  {Gross}}, \bibinfo {author} {\bibfnamefont {T.}~\bibnamefont {Zibold}},
  \bibinfo {author} {\bibfnamefont {E.}~\bibnamefont {Nicklas}}, \bibinfo
  {author} {\bibfnamefont {J.}~\bibnamefont {Esteve}}, \ and\ \bibinfo {author}
  {\bibfnamefont {M.~K.}\ \bibnamefont {Oberthaler}},\ }\href@noop {}
  {\bibfield  {journal} {\bibinfo  {journal} {Nature}\ }\textbf {\bibinfo
  {volume} {464}},\ \bibinfo {pages} {1165} (\bibinfo {year}
  {2010})}\BibitemShut {NoStop}%
\bibitem [{\citenamefont {L{\"u}cke}\ \emph {et~al.}(2011)\citenamefont
  {L{\"u}cke}, \citenamefont {Scherer}, \citenamefont {Kruse}, \citenamefont
  {Pezz{\'e}}, \citenamefont {Deuretzbacher}, \citenamefont {Hyllus},
  \citenamefont {Topic}, \citenamefont {Peise}, \citenamefont {Ertmer},
  \citenamefont {Arlt} \emph {et~al.}}]{lucke2011}%
  \BibitemOpen
  \bibfield  {author} {\bibinfo {author} {\bibfnamefont {B.}~\bibnamefont
  {L{\"u}cke}}, \bibinfo {author} {\bibfnamefont {M.}~\bibnamefont {Scherer}},
  \bibinfo {author} {\bibfnamefont {J.}~\bibnamefont {Kruse}}, \bibinfo
  {author} {\bibfnamefont {L.}~\bibnamefont {Pezz{\'e}}}, \bibinfo {author}
  {\bibfnamefont {F.}~\bibnamefont {Deuretzbacher}}, \bibinfo {author}
  {\bibfnamefont {P.}~\bibnamefont {Hyllus}}, \bibinfo {author} {\bibfnamefont
  {O.}~\bibnamefont {Topic}}, \bibinfo {author} {\bibfnamefont
  {J.}~\bibnamefont {Peise}}, \bibinfo {author} {\bibfnamefont
  {W.}~\bibnamefont {Ertmer}}, \bibinfo {author} {\bibfnamefont
  {J.}~\bibnamefont {Arlt}},  \emph {et~al.},\ }\href@noop {} {\bibfield
  {journal} {\bibinfo  {journal} {Science}\ }\textbf {\bibinfo {volume}
  {334}},\ \bibinfo {pages} {773} (\bibinfo {year} {2011})}\BibitemShut
  {NoStop}%
\bibitem [{\citenamefont {Riedel}\ \emph {et~al.}(2010)\citenamefont {Riedel},
  \citenamefont {B{\"o}hi}, \citenamefont {Li}, \citenamefont {H{\"a}nsch},
  \citenamefont {Sinatra},\ and\ \citenamefont {Treutlein}}]{riedel2010}%
  \BibitemOpen
  \bibfield  {author} {\bibinfo {author} {\bibfnamefont {M.~F.}\ \bibnamefont
  {Riedel}}, \bibinfo {author} {\bibfnamefont {P.}~\bibnamefont {B{\"o}hi}},
  \bibinfo {author} {\bibfnamefont {Y.}~\bibnamefont {Li}}, \bibinfo {author}
  {\bibfnamefont {T.~W.}\ \bibnamefont {H{\"a}nsch}}, \bibinfo {author}
  {\bibfnamefont {A.}~\bibnamefont {Sinatra}}, \ and\ \bibinfo {author}
  {\bibfnamefont {P.}~\bibnamefont {Treutlein}},\ }\href@noop {} {\bibfield
  {journal} {\bibinfo  {journal} {Nature}\ }\textbf {\bibinfo {volume} {464}},\
  \bibinfo {pages} {1170} (\bibinfo {year} {2010})}\BibitemShut {NoStop}%
\bibitem [{\citenamefont {Bloch}\ \emph {et~al.}(2008)\citenamefont {Bloch},
  \citenamefont {Dalibard},\ and\ \citenamefont {Zwerger}}]{RevModPhys.80.885}%
  \BibitemOpen
  \bibfield  {author} {\bibinfo {author} {\bibfnamefont {I.}~\bibnamefont
  {Bloch}}, \bibinfo {author} {\bibfnamefont {J.}~\bibnamefont {Dalibard}}, \
  and\ \bibinfo {author} {\bibfnamefont {W.}~\bibnamefont {Zwerger}},\ }\href
  {\doibase 10.1103/RevModPhys.80.885} {\bibfield  {journal} {\bibinfo
  {journal} {Rev. Mod. Phys.}\ }\textbf {\bibinfo {volume} {80}},\ \bibinfo
  {pages} {885} (\bibinfo {year} {2008})}\BibitemShut {NoStop}%
\bibitem [{\citenamefont {Ockeloen}\ \emph {et~al.}(2013)\citenamefont
  {Ockeloen}, \citenamefont {Schmied}, \citenamefont {Riedel},\ and\
  \citenamefont {Treutlein}}]{ockeloen2013}%
  \BibitemOpen
  \bibfield  {author} {\bibinfo {author} {\bibfnamefont {C.~F.}\ \bibnamefont
  {Ockeloen}}, \bibinfo {author} {\bibfnamefont {R.}~\bibnamefont {Schmied}},
  \bibinfo {author} {\bibfnamefont {M.~F.}\ \bibnamefont {Riedel}}, \ and\
  \bibinfo {author} {\bibfnamefont {P.}~\bibnamefont {Treutlein}},\ }\href@noop
  {} {\bibfield  {journal} {\bibinfo  {journal} {Phys. Rev. Lett.}\ }\textbf
  {\bibinfo {volume} {111}},\ \bibinfo {pages} {143001} (\bibinfo {year}
  {2013})}\BibitemShut {NoStop}%
\bibitem [{\citenamefont {Calarco}\ \emph {et~al.}(2000)\citenamefont
  {Calarco}, \citenamefont {Hinds}, \citenamefont {Jaksch}, \citenamefont
  {Schmiedmayer}, \citenamefont {Cirac},\ and\ \citenamefont
  {Zoller}}]{calarco2000}%
  \BibitemOpen
  \bibfield  {author} {\bibinfo {author} {\bibfnamefont {T.}~\bibnamefont
  {Calarco}}, \bibinfo {author} {\bibfnamefont {E.~A.}\ \bibnamefont {Hinds}},
  \bibinfo {author} {\bibfnamefont {D.}~\bibnamefont {Jaksch}}, \bibinfo
  {author} {\bibfnamefont {J.}~\bibnamefont {Schmiedmayer}}, \bibinfo {author}
  {\bibfnamefont {J.~I.}\ \bibnamefont {Cirac}}, \ and\ \bibinfo {author}
  {\bibfnamefont {P.}~\bibnamefont {Zoller}},\ }\href@noop {} {\bibfield
  {journal} {\bibinfo  {journal} {Phys. Rev. A}\ }\textbf {\bibinfo {volume}
  {61}},\ \bibinfo {pages} {022304} (\bibinfo {year} {2000})}\BibitemShut
  {NoStop}%
\bibitem [{\citenamefont {Kielpinski}\ \emph {et~al.}(2002)\citenamefont
  {Kielpinski}, \citenamefont {Monroe},\ and\ \citenamefont
  {Wineland}}]{kielpinski2002}%
  \BibitemOpen
  \bibfield  {author} {\bibinfo {author} {\bibfnamefont {D.}~\bibnamefont
  {Kielpinski}}, \bibinfo {author} {\bibfnamefont {C.}~\bibnamefont {Monroe}},
  \ and\ \bibinfo {author} {\bibfnamefont {D.~J.}\ \bibnamefont {Wineland}},\
  }\href@noop {} {\bibfield  {journal} {\bibinfo  {journal} {Nature}\ }\textbf
  {\bibinfo {volume} {417}},\ \bibinfo {pages} {709} (\bibinfo {year}
  {2002})}\BibitemShut {NoStop}%
\bibitem [{\citenamefont {van Frank}\ \emph {et~al.}(2016)\citenamefont {van
  Frank}, \citenamefont {Bonneau}, \citenamefont {Schmeidmayer}, \citenamefont
  {Hild}, \citenamefont {Gross}, \citenamefont {Cheneau}, \citenamefont
  {Bloch}, \citenamefont {Pichler}, \citenamefont {Negretti}, \citenamefont
  {Calarco},\ and\ \citenamefont {Montangero}}]{Nature}%
  \BibitemOpen
  \bibfield  {author} {\bibinfo {author} {\bibfnamefont {S.}~\bibnamefont {van
  Frank}}, \bibinfo {author} {\bibfnamefont {M.}~\bibnamefont {Bonneau}},
  \bibinfo {author} {\bibfnamefont {J.}~\bibnamefont {Schmeidmayer}}, \bibinfo
  {author} {\bibfnamefont {S.}~\bibnamefont {Hild}}, \bibinfo {author}
  {\bibfnamefont {C.}~\bibnamefont {Gross}}, \bibinfo {author} {\bibfnamefont
  {M.}~\bibnamefont {Cheneau}}, \bibinfo {author} {\bibfnamefont
  {I.}~\bibnamefont {Bloch}}, \bibinfo {author} {\bibfnamefont
  {T.}~\bibnamefont {Pichler}}, \bibinfo {author} {\bibfnamefont
  {A.}~\bibnamefont {Negretti}}, \bibinfo {author} {\bibfnamefont
  {T.}~\bibnamefont {Calarco}}, \ and\ \bibinfo {author} {\bibfnamefont
  {S.}~\bibnamefont {Montangero}},\ }\href@noop {} {\bibfield  {journal}
  {\bibinfo  {journal} {Sci. Rep.}\ }\textbf {\bibinfo {volume} {6}},\ \bibinfo
  {pages} {1} (\bibinfo {year} {2016})}\BibitemShut {NoStop}%
\bibitem [{\citenamefont {Peirce}\ \emph {et~al.}(1988)\citenamefont {Peirce},
  \citenamefont {Dahleh},\ and\ \citenamefont {Rabitz}}]{peirce1988}%
  \BibitemOpen
  \bibfield  {author} {\bibinfo {author} {\bibfnamefont {A.~P.}\ \bibnamefont
  {Peirce}}, \bibinfo {author} {\bibfnamefont {M.~A.}\ \bibnamefont {Dahleh}},
  \ and\ \bibinfo {author} {\bibfnamefont {H.}~\bibnamefont {Rabitz}},\
  }\href@noop {} {\bibfield  {journal} {\bibinfo  {journal} {Phys. Rev. A}\
  }\textbf {\bibinfo {volume} {37}},\ \bibinfo {pages} {4950} (\bibinfo {year}
  {1988})}\BibitemShut {NoStop}%
\bibitem [{\citenamefont {Koch}\ \emph {et~al.}(2004)\citenamefont {Koch},
  \citenamefont {Palao}, \citenamefont {Kosloff},\ and\ \citenamefont
  {Masnou-Seeuws}}]{koch2004}%
  \BibitemOpen
  \bibfield  {author} {\bibinfo {author} {\bibfnamefont {C.~P.}\ \bibnamefont
  {Koch}}, \bibinfo {author} {\bibfnamefont {J.~P.}\ \bibnamefont {Palao}},
  \bibinfo {author} {\bibfnamefont {R.}~\bibnamefont {Kosloff}}, \ and\
  \bibinfo {author} {\bibfnamefont {F.}~\bibnamefont {Masnou-Seeuws}},\
  }\href@noop {} {\bibfield  {journal} {\bibinfo  {journal} {Phys. Rev. A}\
  }\textbf {\bibinfo {volume} {70}},\ \bibinfo {pages} {013402} (\bibinfo
  {year} {2004})}\BibitemShut {NoStop}%
\bibitem [{\citenamefont {Kirk}(2004)}]{kirk2004optimal}%
  \BibitemOpen
  \bibfield  {author} {\bibinfo {author} {\bibfnamefont {D.~E.}\ \bibnamefont
  {Kirk}},\ }\href@noop {} {\emph {\bibinfo {title} {{Optimal control theory:
  An introduction}}}}\ (\bibinfo  {publisher} {Courier Corporation},\ \bibinfo
  {year} {2004})\BibitemShut {NoStop}%
\bibitem [{\citenamefont {Brif}\ \emph {et~al.}(2010)\citenamefont {Brif},
  \citenamefont {Chakrabarti},\ and\ \citenamefont {Rabitz}}]{brif2010control}%
  \BibitemOpen
  \bibfield  {author} {\bibinfo {author} {\bibfnamefont {C.}~\bibnamefont
  {Brif}}, \bibinfo {author} {\bibfnamefont {R.}~\bibnamefont {Chakrabarti}}, \
  and\ \bibinfo {author} {\bibfnamefont {H.}~\bibnamefont {Rabitz}},\
  }\href@noop {} {\bibfield  {journal} {\bibinfo  {journal} {New J. Phys.}\
  }\textbf {\bibinfo {volume} {12}},\ \bibinfo {pages} {075008} (\bibinfo
  {year} {2010})}\BibitemShut {NoStop}%
\bibitem [{\citenamefont {Mennemann}\ \emph {et~al.}(2015)\citenamefont
  {Mennemann}, \citenamefont {Matthes}, \citenamefont {Weishaupl},\ and\
  \citenamefont {Langen}}]{Mennemann}%
  \BibitemOpen
  \bibfield  {author} {\bibinfo {author} {\bibfnamefont {J.}~\bibnamefont
  {Mennemann}}, \bibinfo {author} {\bibfnamefont {D.}~\bibnamefont {Matthes}},
  \bibinfo {author} {\bibfnamefont {R.}~\bibnamefont {Weishaupl}}, \ and\
  \bibinfo {author} {\bibfnamefont {T.}~\bibnamefont {Langen}},\ }\href@noop {}
  {\bibfield  {journal} {\bibinfo  {journal} {New J. Phys.}\ }\textbf {\bibinfo
  {volume} {17}},\ \bibinfo {pages} {113027} (\bibinfo {year}
  {2015})}\BibitemShut {NoStop}%
\bibitem [{\citenamefont {Hohenester}\ \emph {et~al.}(2007)\citenamefont
  {Hohenester}, \citenamefont {Rekdal}, \citenamefont {Borzi},\ and\
  \citenamefont {Schmiedmayer}}]{Hohenester}%
  \BibitemOpen
  \bibfield  {author} {\bibinfo {author} {\bibfnamefont {U.}~\bibnamefont
  {Hohenester}}, \bibinfo {author} {\bibfnamefont {P.~K.}\ \bibnamefont
  {Rekdal}}, \bibinfo {author} {\bibfnamefont {A.}~\bibnamefont {Borzi}}, \
  and\ \bibinfo {author} {\bibfnamefont {J.}~\bibnamefont {Schmiedmayer}},\
  }\href@noop {} {\bibfield  {journal} {\bibinfo  {journal} {Phys. Rev. A}\
  }\textbf {\bibinfo {volume} {75}},\ \bibinfo {pages} {023602} (\bibinfo
  {year} {2007})}\BibitemShut {NoStop}%
\bibitem [{\citenamefont {Bryson}\ and\ \citenamefont {Ho}(1975)}]{Bryson}%
  \BibitemOpen
  \bibfield  {author} {\bibinfo {author} {\bibfnamefont {A.~E.}\ \bibnamefont
  {Bryson}}\ and\ \bibinfo {author} {\bibfnamefont {Y.-C.}\ \bibnamefont
  {Ho}},\ }\href@noop {} {\emph {\bibinfo {title} {{Applied optimal control:
  Optimization, estimation, and control}}}}\ (\bibinfo  {publisher} {Hemisphere
  Publishing Corporation},\ \bibinfo {year} {1975})\BibitemShut {NoStop}%
\bibitem [{\citenamefont {McShane}(1989)}]{Calculus1989}%
  \BibitemOpen
  \bibfield  {author} {\bibinfo {author} {\bibfnamefont {E.~J.}\ \bibnamefont
  {McShane}},\ }\href@noop {} {\bibfield  {journal} {\bibinfo  {journal} {SIAM
  (Society for Industrial and Applied Mathematics) Journal on Control and
  Optimization}\ }\textbf {\bibinfo {volume} {27}},\ \bibinfo {pages} {916}
  (\bibinfo {year} {1989})}\BibitemShut {NoStop}%
\bibitem [{\citenamefont {Hintermuller}\ \emph {et~al.}(2013)\citenamefont
  {Hintermuller}, \citenamefont {Marahrens}, \citenamefont {Markowich},\ and\
  \citenamefont {Sparber}}]{Hintermuller}%
  \BibitemOpen
  \bibfield  {author} {\bibinfo {author} {\bibfnamefont {M.}~\bibnamefont
  {Hintermuller}}, \bibinfo {author} {\bibfnamefont {D.}~\bibnamefont
  {Marahrens}}, \bibinfo {author} {\bibfnamefont {P.~A.}\ \bibnamefont
  {Markowich}}, \ and\ \bibinfo {author} {\bibfnamefont {C.}~\bibnamefont
  {Sparber}},\ }\href@noop {} {\bibfield  {journal} {\bibinfo  {journal} {SIAM
  J. Control Optim.}\ }\textbf {\bibinfo {volume} {51}},\ \bibinfo {pages}
  {2509} (\bibinfo {year} {2013})}\BibitemShut {NoStop}%
\bibitem [{\citenamefont {Goodman}(2011)}]{JPhysA2011}%
  \BibitemOpen
  \bibfield  {author} {\bibinfo {author} {\bibfnamefont {R.~H.}\ \bibnamefont
  {Goodman}},\ }\href {http://iopscience.iop.org/1751-8121/44/42/425101}
  {\bibfield  {journal} {\bibinfo  {journal} {J. Phys. A: Math. Theor.}\
  }\textbf {\bibinfo {volume} {44}},\ \bibinfo {pages} {425101} (\bibinfo
  {year} {2011})}\BibitemShut {NoStop}%
\bibitem [{\citenamefont {Goodman}\ \emph {et~al.}(2015)\citenamefont
  {Goodman}, \citenamefont {Marzuola},\ and\ \citenamefont
  {Weinstein}}]{gmw_2015}%
  \BibitemOpen
  \bibfield  {author} {\bibinfo {author} {\bibfnamefont {R.~H.}\ \bibnamefont
  {Goodman}}, \bibinfo {author} {\bibfnamefont {J.~L.}\ \bibnamefont
  {Marzuola}}, \ and\ \bibinfo {author} {\bibfnamefont {M.~I.}\ \bibnamefont
  {Weinstein}},\ }\href
  {http://www.aimsciences.org/journals/displayArticlesnew.jsp?paperID=10218}
  {\bibfield  {journal} {\bibinfo  {journal} {Disc. Cont. Dyn. Sys. A}\
  }\textbf {\bibinfo {volume} {35}},\ \bibinfo {pages} {225} (\bibinfo {year}
  {2015})}\BibitemShut {NoStop}%
\bibitem [{\citenamefont {Doria}\ \emph {et~al.}(2011)\citenamefont {Doria},
  \citenamefont {Calarco},\ and\ \citenamefont {Montangero}}]{Doria}%
  \BibitemOpen
  \bibfield  {author} {\bibinfo {author} {\bibfnamefont {P.}~\bibnamefont
  {Doria}}, \bibinfo {author} {\bibfnamefont {T.}~\bibnamefont {Calarco}}, \
  and\ \bibinfo {author} {\bibfnamefont {S.}~\bibnamefont {Montangero}},\
  }\href {\doibase 10.1103/PhysRevLett.106.190501} {\bibfield  {journal}
  {\bibinfo  {journal} {Phys. Rev. Lett.}\ }\textbf {\bibinfo {volume} {106}},\
  \bibinfo {pages} {190501} (\bibinfo {year} {2011})}\BibitemShut {NoStop}%
\bibitem [{\citenamefont {Caneva}\ \emph {et~al.}(2011)\citenamefont {Caneva},
  \citenamefont {Calarco},\ and\ \citenamefont {Montangero}}]{Caneva}%
  \BibitemOpen
  \bibfield  {author} {\bibinfo {author} {\bibfnamefont {T.}~\bibnamefont
  {Caneva}}, \bibinfo {author} {\bibfnamefont {T.}~\bibnamefont {Calarco}}, \
  and\ \bibinfo {author} {\bibfnamefont {S.}~\bibnamefont {Montangero}},\
  }\href {\doibase 10.1103/PhysRevA.84.022326} {\bibfield  {journal} {\bibinfo
  {journal} {Phys. Rev. A}\ }\textbf {\bibinfo {volume} {84}},\ \bibinfo
  {pages} {022326} (\bibinfo {year} {2011})}\BibitemShut {NoStop}%
\bibitem [{\citenamefont {Storn}\ and\ \citenamefont {Price}(1997)}]{Storn}%
  \BibitemOpen
  \bibfield  {author} {\bibinfo {author} {\bibfnamefont {R.}~\bibnamefont
  {Storn}}\ and\ \bibinfo {author} {\bibfnamefont {K.}~\bibnamefont {Price}},\
  }\href@noop {} {\bibfield  {journal} {\bibinfo  {journal} {J. Global Optim.}\
  }\textbf {\bibinfo {volume} {11}},\ \bibinfo {pages} {341} (\bibinfo {year}
  {1997})}\BibitemShut {NoStop}%
\bibitem [{\citenamefont {Viana-Gomes}\ and\ \citenamefont
  {Peres}(2011)}]{Viana-Gomes2011}%
  \BibitemOpen
  \bibfield  {author} {\bibinfo {author} {\bibfnamefont {J.}~\bibnamefont
  {Viana-Gomes}}\ and\ \bibinfo {author} {\bibfnamefont {N.~M.~R.}\
  \bibnamefont {Peres}},\ }\href@noop {} {\bibfield  {journal} {\bibinfo
  {journal} {Eur. J. Phys}\ }\textbf {\bibinfo {volume} {32}},\ \bibinfo
  {pages} {1377} (\bibinfo {year} {2011})}\BibitemShut {NoStop}%
\bibitem [{\citenamefont {S{\o}rensen}\ \emph {et~al.}(2018)\citenamefont
  {S{\o}rensen}, \citenamefont {Aranburu}, \citenamefont {Heinzel},\ and\
  \citenamefont {Sherson}}]{Sorensen}%
  \BibitemOpen
  \bibfield  {author} {\bibinfo {author} {\bibfnamefont {J.~J. W.~H.}\
  \bibnamefont {S{\o}rensen}}, \bibinfo {author} {\bibfnamefont {M.~O.}\
  \bibnamefont {Aranburu}}, \bibinfo {author} {\bibfnamefont {T.}~\bibnamefont
  {Heinzel}}, \ and\ \bibinfo {author} {\bibfnamefont {J.~F.}\ \bibnamefont
  {Sherson}},\ }\href {\doibase 10.1103/PhysRevA.98.022119} {\bibfield
  {journal} {\bibinfo  {journal} {Phys. Rev. A}\ }\textbf {\bibinfo {volume}
  {98}},\ \bibinfo {pages} {022119} (\bibinfo {year} {2018})}\BibitemShut
  {NoStop}%
\bibitem [{\citenamefont {Boyd}\ and\ \citenamefont
  {Vandenberghe}(2004)}]{BoydV}%
  \BibitemOpen
  \bibfield  {author} {\bibinfo {author} {\bibfnamefont {S.}~\bibnamefont
  {Boyd}}\ and\ \bibinfo {author} {\bibfnamefont {L.}~\bibnamefont
  {Vandenberghe}},\ }\href@noop {} {\emph {\bibinfo {title} {Convex
  optimization}}},\ Vol.~\bibinfo {volume} {1}\ (\bibinfo  {publisher}
  {Cambridge University Press},\ \bibinfo {address} {Cambridge},\ \bibinfo
  {year} {2004})\BibitemShut {NoStop}%
\bibitem [{\citenamefont {Price}\ \emph {et~al.}(2018)\citenamefont {Price},
  \citenamefont {Storn},\ and\ \citenamefont {Lampinen}}]{DiffEvoBook}%
  \BibitemOpen
  \bibfield  {author} {\bibinfo {author} {\bibfnamefont {K.}~\bibnamefont
  {Price}}, \bibinfo {author} {\bibfnamefont {R.}~\bibnamefont {Storn}}, \ and\
  \bibinfo {author} {\bibfnamefont {J.}~\bibnamefont {Lampinen}},\ }\href@noop
  {} {\emph {\bibinfo {title} {Differential {E}volution: A practical approach
  to global optimization}}}\ (\bibinfo  {publisher} {Springer-Verlag Berlin and
  Heidelberg GmbH \& Co. KG},\ \bibinfo {address} {Berlin},\ \bibinfo {year}
  {2018})\BibitemShut {NoStop}%
\bibitem [{\citenamefont {Trefethen}(2000)}]{Trefethen}%
  \BibitemOpen
  \bibfield  {author} {\bibinfo {author} {\bibfnamefont {L.~N.}\ \bibnamefont
  {Trefethen}},\ }\href@noop {} {\emph {\bibinfo {title} {Spectral methods in
  MATLAB}}}\ (\bibinfo  {publisher} {SIAM},\ \bibinfo {address} {New York,
  NY},\ \bibinfo {year} {2000})\BibitemShut {NoStop}%
\bibitem [{\citenamefont {von Winckel}\ and\ \citenamefont
  {Borzi}(2008)}]{vonWinckel}%
  \BibitemOpen
  \bibfield  {author} {\bibinfo {author} {\bibfnamefont {G.}~\bibnamefont {von
  Winckel}}\ and\ \bibinfo {author} {\bibfnamefont {A.}~\bibnamefont {Borzi}},\
  }\href@noop {} {\bibfield  {journal} {\bibinfo  {journal} {Inverse Probl.}\
  }\textbf {\bibinfo {volume} {24}},\ \bibinfo {pages} {034007} (\bibinfo
  {year} {2008})}\BibitemShut {NoStop}%
\end{thebibliography}%


%
\end{document}